\documentclass[twocolumn]{aastex63}

\usepackage{lineno}
\usepackage{comment}
\usepackage[figure,figure*]{hypcap}
\usepackage{xcolor}
\usepackage{xspace}
\usepackage{graphicx,natbib,placeins,amsmath}

\usepackage{multirow}
\usepackage{amssymb}
\usepackage{enumitem}
\usepackage{url}
\usepackage{enumitem}
\usepackage{soul}   

\DeclareUnicodeCharacter{2009}{\,}

\newcommand{\Teff} {\ensuremath{T_{\rm eff}}\xspace}

\newcommand{\Av} {\ensuremath{A_{V}}\xspace}
\newcommand{\Rv} {\ensuremath{R_{V}}\xspace}
\newcommand{\Ha} {\ensuremath{{\rm H}\alpha}\xspace}

\newcommand{\Msun}{\ensuremath{M_{\odot}}\xspace}

\newcommand{\Mstar}{\ensuremath{M_{\star}}\xspace}
\newcommand{\logMstar}{\ensuremath{\log M_{\star}}\xspace}
\newcommand{\Lsun}{\ensuremath{L_{\odot}}\xspace}

\newcommand{\Lstar}{\ensuremath{L_{\star}}\xspace}
\newcommand{\logLstar}{\ensuremath{\log L_{\star}}\xspace}

\newcommand{\Rstar}{\ensuremath{R_{\star}}\xspace}

\newcommand{\Lacc}{\ensuremath{L_{acc}}\xspace}
\newcommand{\logLacc}{\ensuremath{\log L_{acc}}\xspace}

\newcommand{\dMacc}{\ensuremath{\dot{M}_{acc}}\xspace}
\newcommand{\logdMacc}{\ensuremath{\log\dot{M}_{acc}}\xspace}
\shorttitle{A Bayesian multicolor study of stellar parameters in the ONC}
\shortauthors{Strampelli et al.}

\begin{document}

\title{HST Survey of the Orion Nebula Cluster in ACS/Visible and WFC3/IR Bands. IV. A Bayesian multi-wavelength study of stellar parameters in the ONC}

\correspondingauthor{Giovanni M. Strampelli}
\email{gstrampelli@ucsb.edu}

\author[0000-0002-1652-420X]{Giovanni M. Strampelli}
\affiliation{University of California Santa Barbara, Santa Barbara, CA 93106, USA}

\author[0000-0002-9573-3199]{Massimo Robberto}
\affiliation{Johns Hopkins University, 3400 N. Charles Street, Baltimore, MD 21218, USA}
\affiliation{Space Telescope Science Institute, 3700 San Martin Dr, Baltimore, MD 21218, USA}

\author{Laurent Pueyo}
\affiliation{Space Telescope Science Institute, 3700 San Martin Dr, Baltimore, MD 21218, USA}

\author{Mario Gennaro}
\affiliation{Space Telescope Science Institute, 3700 San Martin Dr, Baltimore, MD 21218, USA}

\author{Carlo F. Manara}
\affiliation{European Southern Observatory, Garching bei München, Germany.}

\author{Elena Sabbi}
\affiliation{Space Telescope Science Institute, 3700 San Martin Dr, Baltimore, MD 21218, USA}

\author{Antonio Aparicio}
\affiliation{Department of Astrophysics, University of La Laguna, Av. Astrofísico Francisco Sánchez, 38200 San Cristóbal de La Laguna, Tenerife, Canary Islands, Spain}
\affiliation{Instituto de Astrofísica de Canarias, C. Vía Láctea, 38200, San Cristóbal de La Laguna, Tenerife, Canary Islands, Spain}


\begin{abstract}
We have performed a comprehensive study of the Orion Nebula Cluster (ONC) combining the photometric data obtained by the two \textit{HST} Treasury programs that targeted this region.
To consistently analyze the rich dataset obtained in a wide variety of filters, we adopted a Bayesian approach to fit the Spectral Energy Distribution of the sources, deriving mass, age, extinction, distance, and accretion for each source in the region. 
The three dimensional study of mass distribution for bona-fide cluster members shows that  mass segregation in the ONC extends to sub-solar masses, while the age distribution strongly supports the idea  that  star  formation  in  the ONC is best described by a major episode of star formation that happened $\sim 1$ Myr ago. 
For masses $\gtrsim 0.1$ \Msun, our derived empirical initial mass function (IMF) is in good agreement with a Chabrier system IMF.
Both the accretion luminosity (\Lacc) and mass accretion rates (\dMacc) are best described by broken power-law relations.
This suggests that for the majority of young circumstellar disks in this cluster the excess emission may be dominated by X-ray-driven photoevaporation by the central star rather than external photoevaporation. If this is the case, the slopes of the power-law relations may be largely determined by the initial conditions set at the onset of the star formation process, which may be quite similar between regions that eventually form clusters of different sizes.
\end{abstract}


\section{Introduction}

\begin{table*}[!t]
    \centering
    \footnotesize
    \begin{tabular}{ccccccc}
    \hline
    \hline
    FILTER      &   PHOTPLAM    &   PHOTFLAM    & Ground Equivalent & Exposure & Integration time &  Zero Point	\\   &   \AA       &   erg cm$^{-2}$ s$^{-1}$ \AA$^{-1}$  &  &  & (s) & Vega mag	 \\
    \hline
    F435W       &	4329.2      &	3.113e-19   & Johnson B & 6 & 420 &	25.776 \\
    F555W	    &   5360.2	    &   1.948e-19	& Johnson V & 9 & 385 & 25.722 \\
    F658N	    &   6584.0	    &   1.977e-18	& Broad \Ha & 1 & 340 & 22.378 \\
    F775W	    &   7693.9	    &   9.928e-20	& Sloan i & 8 &  385 & 25.275 \\
    F850LP	    &   9034.6	    &   1.503e-19   & Sloan z & 7 &  385 & 24.345 \\
    \hline
    \end{tabular}
    \caption{ACS visits strategy and photometric system from \cite{Robberto2013}. The zero points for each filter are obtained from the STScI \textit{ACS Zeropoints Calculator}.}
    \label{tab:ZPT}
\end{table*}

The Orion Nebula Cluster (ONC) represents one of the richest \citep[$\sim 2000$ members in the inner 2 pc radius;][]{2019ApJ...884....6M} and youngest \citep[$\sim 2$ Myr; ][]{Jeffries2011,Reggiani2011,Jerabkova2019} star-forming regions (SFRs) within 2 kpc \citep[$\sim 402$ pc; ][]{Kuhn2019}  from the sun \citep{Lada2003,Portegies2010}. Due to its proximity and modest foreground extinction \cite[$\Av\sim 1$][]{Scandariato2011}, the ONC 
has been thoroughly studied at visible and near-infrared wavelengths  and is therefore regarded as an ideal laboratory where to investigate, down to very low mass objects,  critical aspects of star and planetary formation, such as the initial mass function, mass segregation and early dynamical evolution,  radiative feedback and protoplanetary disk photoevaporation and evolution.  


In this paper, we combine for the first time the data coming from the two \textit{HST} Treasury programs (GO-10246 and GO-13826, P.I. M. Robberto) that targeted the ONC. 
Taking advantage of the extremely accurate photometry provided by the HST in a wide selection of filters, we adopt a Bayesian approach with Markov Chain Monte Carlo (MCMC) strategy to estimate the main properties of each star in the cluster (e.g. mass, extinction and age, distance and accretion) according to the BT-Settl family of isochrones. A Bayesian approach allows incorporating as priors other available information e.g. on the source effective temperature, distance, reddening, membership, etc. obtaining a most accurate and consistent estimate of the stellar parameters.


Our first aim is to assess the viability of this technique, considering that our dataset is largely based on archival data, taken at different epochs, and the number of filters available for each source is not uniform across the sample. Concentrating on the most reliable sources, we then perform a deeper analysis of the star formation and evolution history of the Orion Nebula Cluster.

The paper is organized as follows: the observations are presented in $\S$\ref{sec:Observation and Data Reduction}, including a new ACS photometric catalog based on the most recent instrument calibration data.
In $\S$\ref{sec:Analysis} we introduce our Bayesian code to perform Spectral Energy Distribution (SED) fitting. In Section $\S$\ref{sec:Results} we present our final catalog of bonafide cluster members, the derived stellar parameters, and the mass accretion luminosities and rates. 
In Section $\S$\ref{sec:Discussion} we discuss the implication of our findings.
Our conclusions are summarized in $\S$\ref{sec:Conclusion}. 

\section{Datasets and new ACS photometry}
\label{sec:Observation and Data Reduction}
\begin{figure*}[!t]
    \centering
    \includegraphics[width=0.30\textwidth]{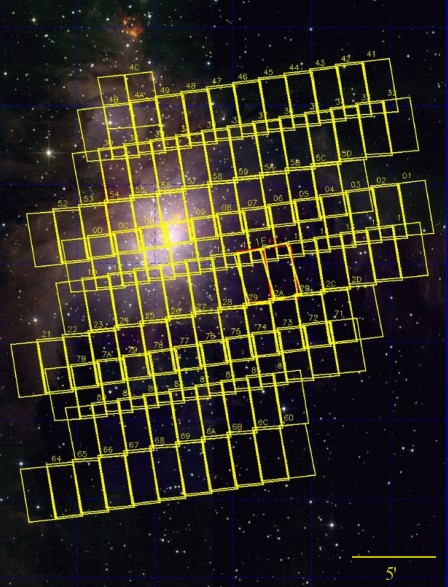}
    \includegraphics[width=0.30\textwidth]{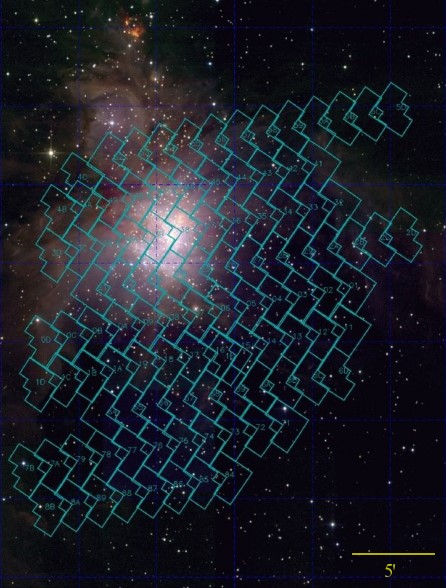}\\
    \includegraphics[width=0.30\textwidth]{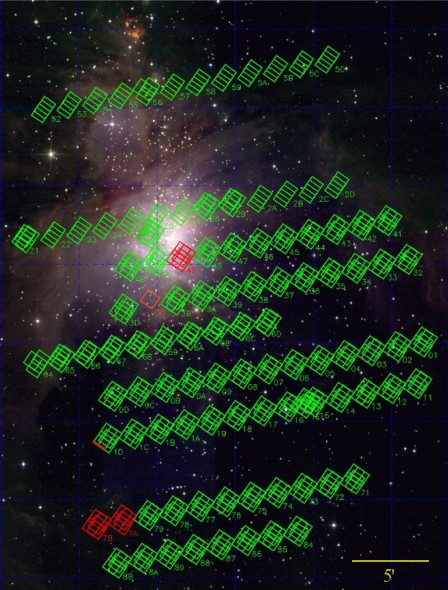}
    \includegraphics[width=0.30\textwidth]{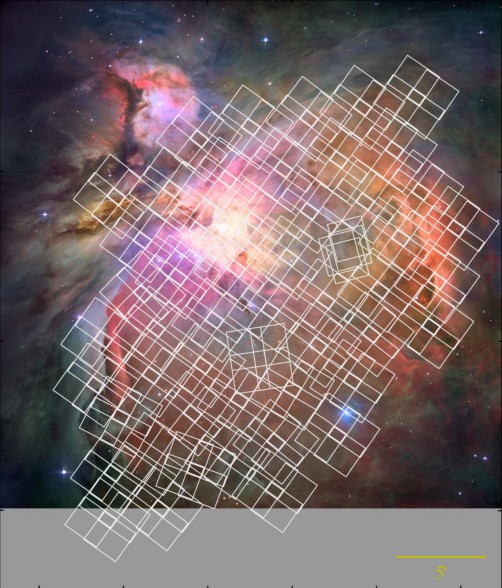}
    \caption{Total area coverage of the different instruments adopted in this work. From top to bottom (left to right): ACS, WFPC2, NICMOS \cite[superimposed to a color-composite JHK image of the Orion Nebula from 2MASS;][]{Robberto2013} and WFC3 \cite[superimposed to a mosaic obtained from both space and ground telescopes data;][]{Robberto2020}.
    For reference, the ACS mosaic covers $\sim 627$ arcmin$^2$, while WFC3-IR covers $\sim 486$ arcmin$^2$.}
    \label{fig:FOW}
\end{figure*}

The HST program GO-10246, executed between October 2004 and March 2005, observed the ONC with the F435W, F555W, F658N, F775W and F850LP filters of ACS/WFC, the F336W, F439W, F656N, and F814W of WFPC2, and F110W and F160W of  NICMOS  \citep{Robberto2013}. The richest dataset is the one provided by ACS, which includes 3399  unique sources, most of them observed two or more times in different \textit{HST} visits, separated by time intervals ranging from a few hours to several months. \cite{Robberto2013} lists all individual detection, a grand total of 8185 entries each with multi-color observations.
In 2015 the HST program GO-13826 returned to the ONC with the F130N and F139M of WFC3  \citep{Robberto2020}. As shown by Figure \ref{fig:FOW}, the areal coverage is similar for the ACS, WFPC2, and WFC3 data, while the NICMOS observations sampled about 1/4 of the field due to the small size of the detector.

In Table \ref{tab:Filter detection} we show the number of sources detected in each filter. We will use this dataset as our reference catalog. 
\begin{table*}[!t]
    \centering
    \begin{tabular}{ccccccccccccc}
        \hline
        \hline
        F435W & F555W & F658N & F775W & F850LP & F130N & F139M & F110W & F160W & F336W & F439W & F656N & F814W\\
        \hline
        1264  &  1525 & 1316 & 2466 & 2725 & 2728 & 2897 & 425 & 518 & 549 & 627 & 1018 & 1365\\
        \hline
    \end{tabular}
    \caption{Number of sources detected in each filter.}
    \label{tab:Filter detection}
\end{table*}

Given both the advances in the ACS calibration and the possibility of 
obtaining more accurate data on close pairs, we reanalyzed the ACS dataset to derive updated aperture photometry. 
In Table \ref{tab:ZPT} we list the main parameters of the photometric system for the five ACS bandpasses including the most recent assessment of the zero points at the epoch of observations,  as provided by the \textit{ACS zero points Calculator}\footnote{\url{https://acszeropoints.stsci.edu/}}. Compared to the original zero points implemented by the \cite{Robberto2013}, the  new values show a difference between $-0.002$ and $0.019$ magnitudes, with  the smallest difference in the F555W filter and the highest one for  both the F775W and F850LP filters. 

Aperture photometry was obtained using the \textit{aperture photometry} package from 
the \textit{Stra}KLIP pipeline \citep{Strampelli2022}. This tool allows the detection and subtraction of the signal from sources very close to the primary target to perform accurate photometry on both components. In this paper we shall focus on the main population of isolated stars, leaving the analysis of close pairs to a future paper.

The individual measures obtained in different visits were finally averaged vetting and rejecting those with high uncertainty due to the residual detector defects or the presence of cosmic rays in the immediate vicinity. 

\begin{figure}[!t]
    \centering
    \includegraphics[width=0.45\textwidth]{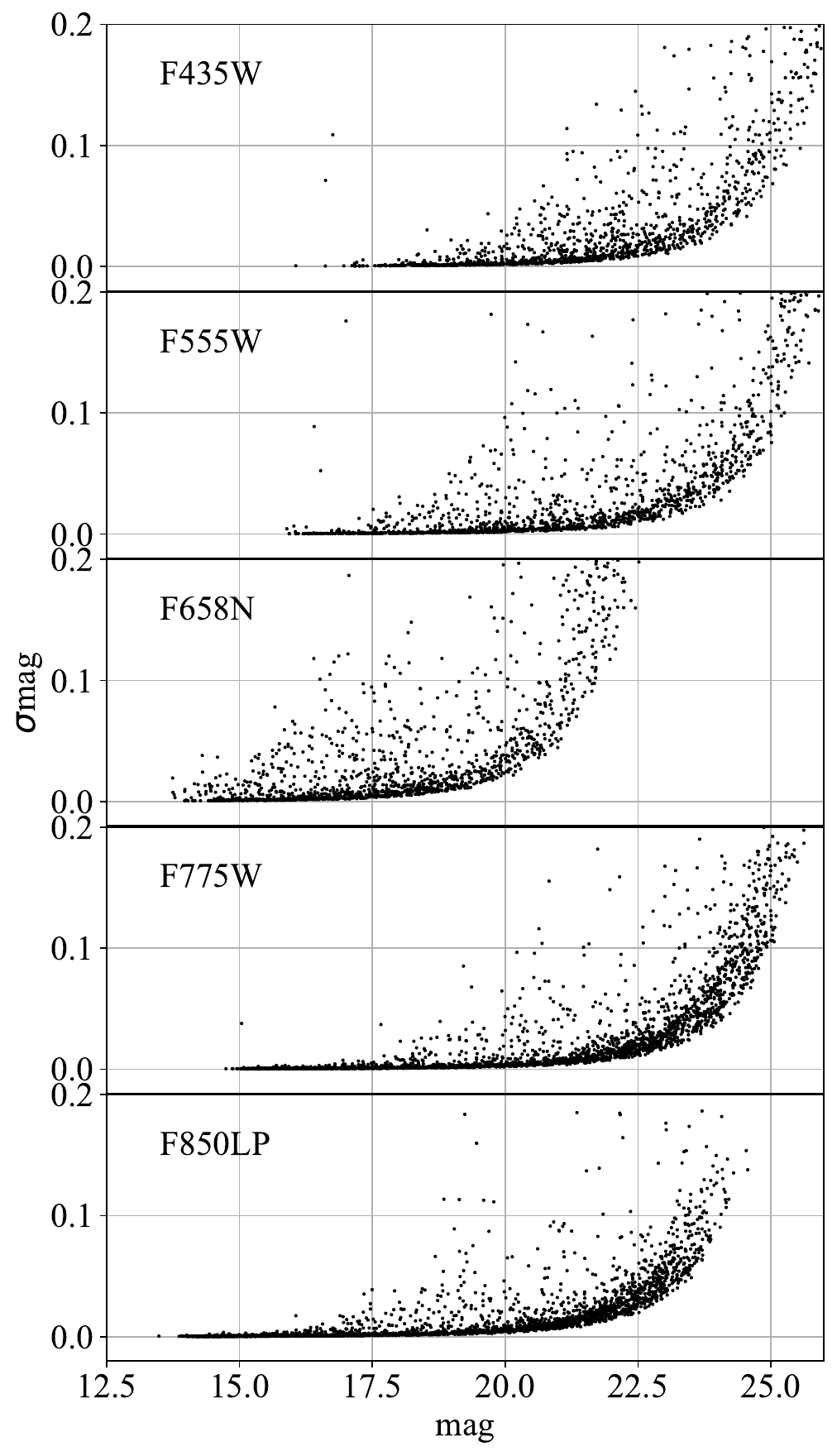}
    \caption{Photometric errors as a function of magnitude for the five ACS filters.}
    \label{fig:errors_plot}
\end{figure}

In Figure \ref{fig:errors_plot} we plot our final averaged magnitudes with their estimated uncertainties for all sources in the catalog in the five ACS filters. Saturation starts at about m$_{435}$ = 16, m$_{555}$ = 15.75, m$_{658}$ = 12.25, m$_{775}$ = 15.25 and m$_{850}$ = 14.25, corresponding respectively to 0.7, 0.5, 1.2, 0.14 and 0.13 \Msun,  assuming the BT-Settl 1M~yr isochrone at 400~pc, without extinction or accretion. 
At the other extreme, the $5\sigma$ sensitivity limits ($\sigma_{mag} \simeq$ 0.2) are at m$_{435}$ = 23.77, m$_{555}$ = 24.07, m$_{658}$ = 20.15, m$_{775}$ = 23.84 and m$_{850}$ = 23.01, roughly corresponding to 0.03, 0.02, 0.04, 0.005 and  0.003\Msun, under the same assumptions. Table \ref{Tab:mass_thersolds} provides the saturation/sensitivity limits as a function of extinction and detectable mass. All magnitudes are in the Vega system.  Apart from ACS, for the other instruments and passbands we directly adopted the catalogs presented by \cite{Robberto2013, Robberto2020}.

\begin{deluxetable*}{lccccccccccccccc}
\tabletypesize{\footnotesize}
\setlength{\tabcolsep}{0.25pt}
\tablecaption{\textit{Sample of the photometric data available for all the sources included in this work.}
\label{Tab:phot}}
\tablehead{
\colhead{id} &
\colhead{spx435} & \colhead{bpx435} & \colhead{m435}  &
\colhead{spx555} & \colhead{bpx555} & \colhead{m555}  &
\colhead{spx658} & \colhead{bpx658} & \colhead{m658}  &
\colhead{spx775} & \colhead{bpx775} & \colhead{m775}  &
\colhead{spx850} & \colhead{bpx850} & \colhead{m850}  \\
\colhead{--}   &
\colhead{(--)}   & \colhead{(--)}  & \colhead{(mag)} &
\colhead{(--)}   & \colhead{(--)}  & \colhead{(mag)} &
\colhead{(--)}   & \colhead{(--)}  & \colhead{(mag)} &
\colhead{(--)}   & \colhead{(--)}  & \colhead{(mag)} &
\colhead{(--)}   & \colhead{(--)}  & \colhead{(mag)}  
}
\startdata
1        &       &       &        &    0   &   2    & $24.591^{+0.106}_{-0.106}$ &       &       &        & 0      & 2      & $21.836^{+0.013}_{-0.013}$ & 0      & 2      & $20.827^{+0.011}_{-0.011}$ \\
 8        & 0      & 4      & $21.583^{+0.008}_{-0.008}$ & 0      & 4      & $19.725^{+0.003}_{-0.003}$ &  0      & 4      & $17.609^{+0.007}_{-0.007}$ & 20     & 4      & $16.387^{+0.001}_{-0.001}$ & 18     & 4      & $15.274^{+0.001}_{-0.001}$ \\
 ...\\
\enddata
\tablecomments{\textit{Table \ref{Tab:phot} is published in its entirety in the machine-readable format. A portion is shown here for guidance regarding its form and content. Note that in the published machine-readable version of this table, the errors will occupy a different column.}}
\end{deluxetable*}

Since most observations were not carried out simultaneously, variability represents a source of uncertainty. According to \cite{Herbst2002}, about half of the stars brighter than $I \simeq 16$ show peak-to-peak variations of $\sim0.2$~mag, or more. While this is typically not the range of magnitudes probed by our deeper HST observations, it suggests that variability randomly affects the quality of the fit. In principle, systematic uncertainties could be reduced using colors combining measures obtained with the same instrument within the same HST orbit, i.e. within about 45~minutes. Testing this approach, we found that in most cases having fewer data points with uncertainties added in quadrature precludes achieving any significant gain compared to the case where the SED fit is performed using single magnitudes. We have therefore carried out our analysis ignoring variability, at least initially, returning later to the sources that may not pass our tests on the quality of the fit because of spurious photometric data for a specific epoch of observation.
Table \ref{Tab:phot} shows a sample of our final photometric catalog for all 3399 sources, fully available in electronic format. The table is organized as follows: the first column reports the ID as provided from the \textit{Stra}KLIP pipeline for cross-identification, while the following columns report the average number of saturated pixels, number of identified bad pixels, magnitude, and uncertainty for each filter.

\section{Analysis}
\label{sec:Analysis}
\subsection{Bayesian analysis}
\label{sec:Bayesian analysis}
For our analysis we assume that the main unknown variables that characterize each source are mass, extinction, age, distance, and the accretion parameter ($SP_{acc}$), representing the fraction of stellar bolometric luminosity that can be attributed to accretion.
. To determine them, we perform Spectral Energy Distribution (SED) fitting comparing the observed photometry to synthetic photometry derived from theoretical models of the BT-Settl family (see Section \ref{Sec:Models}). In order to obtain a probability distribution for each  parameter, our fitting procedure adopts a Bayesian approach with an MCMC algorithm, represented schematically in Figure \ref{fig:ONC_mcmc} and based on the following main steps:
 
\begin{figure*}[!t]
    \centering
    \includegraphics[width=0.95\textwidth]{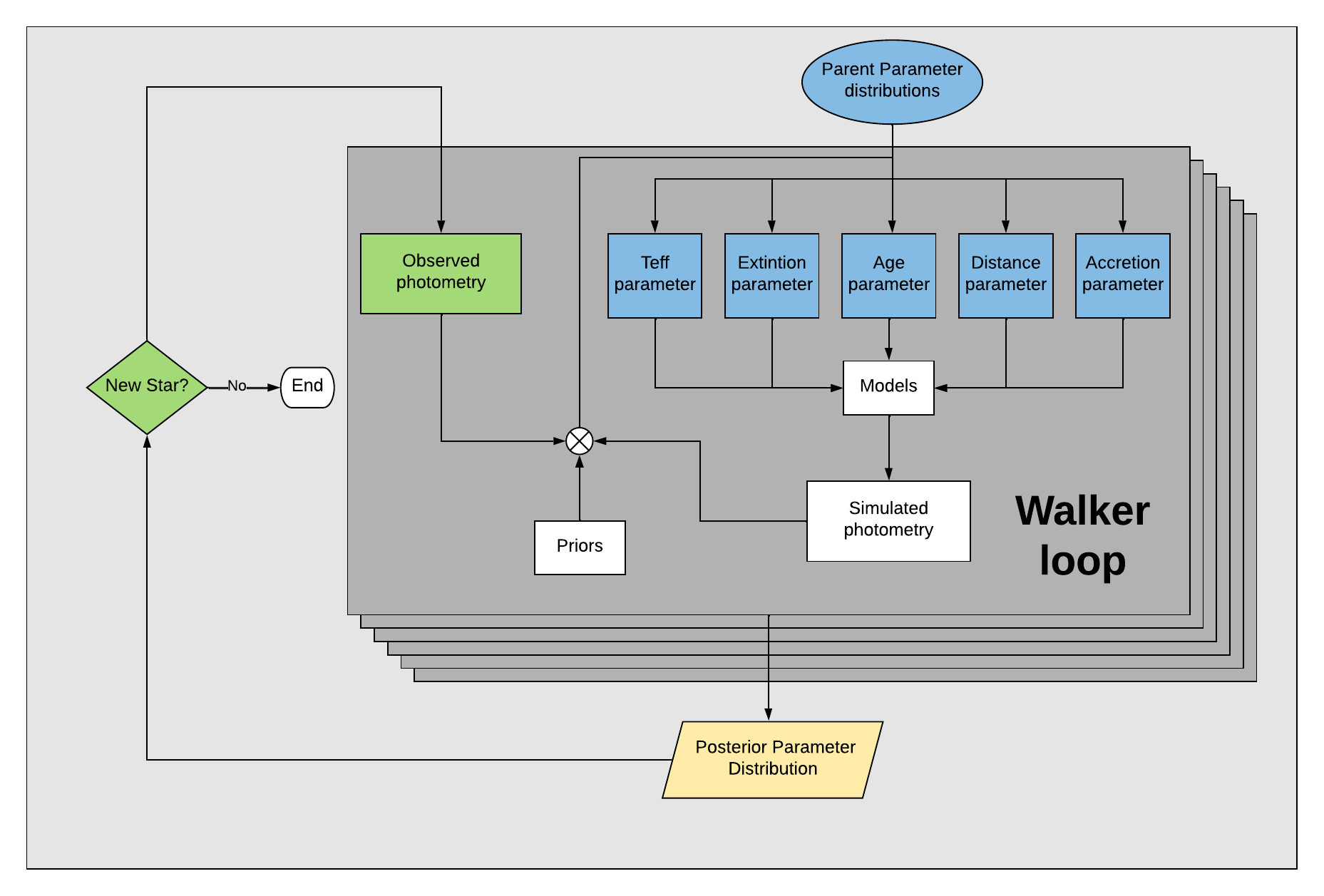} 
    \caption{Flowchart for the ONC sources parameter estimations. The light grey area represents the whole MCMC run while the dark one represents each single walker loop. The ensemble of all walker loops represents the simulation for one single star. The green colored blocks represent fixed ingredients for the simulation while in blue are shown the fitting parameters and in white the key steps of the Bayesian approach. The yellow block shows the stored final output of each iteration. Note that all stars are treated completely independently until the endpoint.
    }
    \label{fig:ONC_mcmc}
\end{figure*}

\begin{table}[!t]
    \caption{Range of values explored for each parameter during our MCMC analysis. From left to right: the stellar mass in Solar units, visible extinction, stellar age in Myr, the rescale factor for the slab model we use to model accretion, and the parallax (corresponding to an interval between 167 pc and 50 Kpc). }
    \footnotesize
    \begin{center}
    \begin{tabular}{cccccc}
    \hline
    \hline
    log$_{10}$(Mass)& log$_{10}$(A$_V$) & log$_{10}$(Age) &  log$_{10}$(SP$_{acc}$) & Parallax & \\
    \hline
    [-2, 1]         &       [-1, 2]     &       [0, 4]    &             [-5, 1]     &  [0.02,6]&\\
    \hline
    \end{tabular}
    \label{tab:par_space}
    \end{center}
\end{table}

\begin{itemize}
    \item at the start of the MCMC run, five parent distributions are generated uniformly spanning the parameter space presented in Table \ref{tab:par_space}.
    
    \item for each star, a user's defined number of \textit{walkers} is generated, limited in this work to a maximum of 100. Then the following loop is iterated for each walker (referred to as the "walker loop" in Figure \ref{fig:ONC_mcmc}):
    \begin{itemize}
        \item  from each distribution discussed above (see also Table \ref{tab:par_space}), a random value is extracted to provide the initial state for the walker in that variable.
        
        \item The \textit{observed magnitudes} of the star under consideration are pulled from the catalog and parsed to the fitting routine as fixed ingredients for the simulation. If one or more filters show sign of saturation, they are discarded from the fit. 
       
        \item The five parameters randomly extracted from their parent distributions are combined with the models (see Section \ref{Sec:Models}) to obtain synthetic magnitudes.

        \item The synthetic and observed magnitudes are iteratively compared to find the set of parameters that better reproduces the observations. This optimization requires: 1) priors on the fitting parameters (see Section \ref{sec:Priors}), 2) a likelihood function to compare observations and simulations (see Section \ref{sec:Likelihood}), 3) an algorithm for sampling the posterior distribution of the fit parameters (see Section \ref{sec:MCMC});
       
        \item The walkers are let to evolve until convergence is reached or a maximum number of steps is executed (2000 in our case). The resulting posterior distribution of the parameters is saved for analysis and the routine moves to the next source;
    \end{itemize}
\end{itemize}
In the following subsections, we detail the main steps of the procedure.

\subsubsection{Models}
\label{Sec:Models}
For our analysis, we use the evolutionary models and spectra of the BT-Settl family (AGSS2009 models using the \cite{2009ARA&A..47..481A} solar abundances)
\footnote{downloaded from \url{http://svo2.cab.inta-csic.es/theory/newov2/}}, spanning the age range between 1 Myr to 10 Gyr and masses from 0.005 \Msun to 1.4~\Msun. These values encompass the range of masses and ages appropriate for our (saturation limited) ONC sample and non-cluster contaminants, based on the Besançon model (see below). 
The fitting routine requires a uniform grid of synthetic photometry in our filters, which we have created using Synphot \citep{synphot2018} and stsynphot \citep{stsynphot2020}. In order to estimate the possible presence of accretion luminosity and  derive mass accretion rates we use a slab model described by \cite{Manara2012} where the combination of optically thin emission generated in the preshock region and optically thick emission generated by the heated photosphere is reproduced by combining a Cloudy spectrum for a standards HII region with an 8000 K black body, respectively contributing about 1/4, and 3/4 of the total accretion luminosity \citep{Gullbring2000}. For the far-UV part of the accretion spectrum ($\lambda \lesssim 3000$ \AA), following \cite{France2011}  we prescribe a linear decrease of flux at wavelengths shorter than the Balmer jump.

To redden the synthetic photometry, we adopt the \cite{Cardelli1989} reddening law parameterized with  $\Rv $= 3.1. 

\subsubsection{Likelihood}
\label{sec:Likelihood}
To compare observations and simulations we use a likelihood defined as:
\begin{equation}
    \mathcal{L}(obs | \theta)=\prod^{N_{mag}}_{i=0} e^{-0.5 \times\big(\frac{mag_{i,obs}-mag_{i,mod}(\theta)}{emag_{i,obs}}\big)^2}
\end{equation}
where \textit{i} runs through the different magnitudes available for each star and the labels \textit{obs} and \textit{mod} represent the observed and model magnitudes (mag) with their uncertainties (emag). The parameter $\theta$ instead represents the whole set of fitting parameters corresponding to the different \textit{mag$_{i,mod}$}.

\subsubsection{Priors}
\label{sec:Priors}
We divide our sources in two classes of objects, each with its own specific priors:
\begin{itemize}
    \item Class A: sources having  both a spectral type from  \cite{Hillenbrand1997} or \cite{Hillenbrand2013}  and a distance estimate from Gaia DR3 \citep{GaiaDR3}. This class has strong priors that allow pinpointing with high accuracy both luminosity and mass. In particular:
    \begin{itemize}
        \item for each star, a prior on the mass is derived from the \cite{Hillenbrand2013} temperature when available, or converting the spectral type from \cite{Hillenbrand1997} to effective temperatures using the \cite{Luhman2003} relation. The prior is assumed to be Gaussian, centered on the mass derived from the effective temperature using the model isochrones and with standard deviation given by the mass spread corresponding to two spectroscopic sub-classes \cite[i.e. the typical uncertainty of the spectral type reported by][]{Hillenbrand2013}. 
        \item similarly, the prior on the distance is a Gaussian centered on the reported parallax from Gaia DR3 \citep{GaiaDR3} and standard deviation given  by the  corresponding parallax uncertainty
    \end{itemize}
    
    \item Class B: these sources, typically fainter, miss either an independent estimate of the spectral type, a Gaia parallax, or both. This is a less robust sample due to the much larger uncertainties on the temperature and/or distance. Still, also for this class, we can define priors on the basis of the results for Class A as follows:
        \begin{itemize}
            \item for the mass prior, if the source has no independent spectral type we start combining the overall mass distribution determined for Class A objects. 
            Since this sample is heavily biased toward the ONC, we complement it by adding the distribution of masses obtained from the Besançon model of stellar population synthesis of the Galaxy in the direction of the ONC \footnote{\url{https://model.obs-besancon.fr/modele_discl.php}}. 
            \item for the distance prior, if the distance is not provided in Gaia DR3, we take a similar approach to the one adopted for the mass. We combine the distributions of parallaxes obtained for Class A objects with the distribution of distance values obtained from the Besançon model (converted to parallax). 
        \end{itemize}

        Given the uncertainty on the membership, ONC or Galaxy, for Class B objects we also define two other priors: one on the age and anoher on the extinction.
        \begin{itemize}
            \item to create the prior on the age, we start combining the distributions obtained for Class A objects and the distribution of age values obtained from the Besançon model. 
            \item for the prior on the extinction, we take only the distributions obtained for Class A objects. We do not adopt the extinction distribution provided by the Besançon model as it does not account for the extinction caused by the OMC-1 molecular cloud, which can be both very large and non-uniform over the ONC field.
        \end{itemize}
\end{itemize}

To obtain the final prior probability distribution (i.e.;  the probability distribution that would link the a priori knowledge about a variable before evidence is taken into account) we applied a Gaussian kernel density  estimator\footnote{\url{https://scikit-learn.org/stable/modules/generated/sklearn.neighbors.KernelDensity.html}} to all distributions for both Class A and B.  From now on we will refer to these  resulting probability density functions (PDFs) as the priors, for short.

\subsubsection{MCMC and posterior probabilities}
\label{sec:MCMC}
To derive the distributions of parameters that best describe our observations we use \textit{emcee}, the \cite{Foreman2016} Python implementation of the \cite{Goodman2010} affine-invariant sampler. For each given set of parameters $\theta$, we evaluate the corresponding value of the magnitudes from the models and then we use the likelihood and prior discussed above to obtain the posterior probability distribution for $\theta$, ensuring that the individual chains  converge and are thinned to retain uncorrelated samples. Basically, this approach consists in picking different points from the sample, in every n-th step. As we are dividing these points from the overall Markov chain, the dependence becomes smaller and we can achieve a mostly independent sample.

The following convergence criteria are checked every 100 iterations:
\begin{itemize}
    \item the chain must be longer than 100 times the estimated auto-correlation time, $\tau_{f}$, i.e. the number of steps needed before the chain loose information about its starting properties;
    \item the estimate of $\tau_{f}$ has changed by less than $1\%$ since the previous check.
\end{itemize}
As long as these criteria are met, the routine keeps running until a total of 2000 post-convergence iterations are performed, or the maximum number of steps is elapsed. Then the routine exits the iteration loop and saves the posterior distribution of the parameters to a file before moving to the next source, till the end of the catalog.

\subsubsection{Bayesian Analysis Products}
\label{sec:Bayesian Analysis Products}
Starting from a catalog of 3399 unique sources, we perform Bayesian SED fitting using MCMC algorithm to evaluate the star's principal parameters, i.e., mass, extinction, age, distance, and the $SP_{acc}$ (see Section \ref{sec:Bayesian analysis}).

\figsetstart
\figsetnum{1}
\figsettitle{ONC SED fitting}
\label{fig:corners}

\figsetend

\begin{figure*}
\label{fig:corners}
\includegraphics[width=0.99\textwidth]{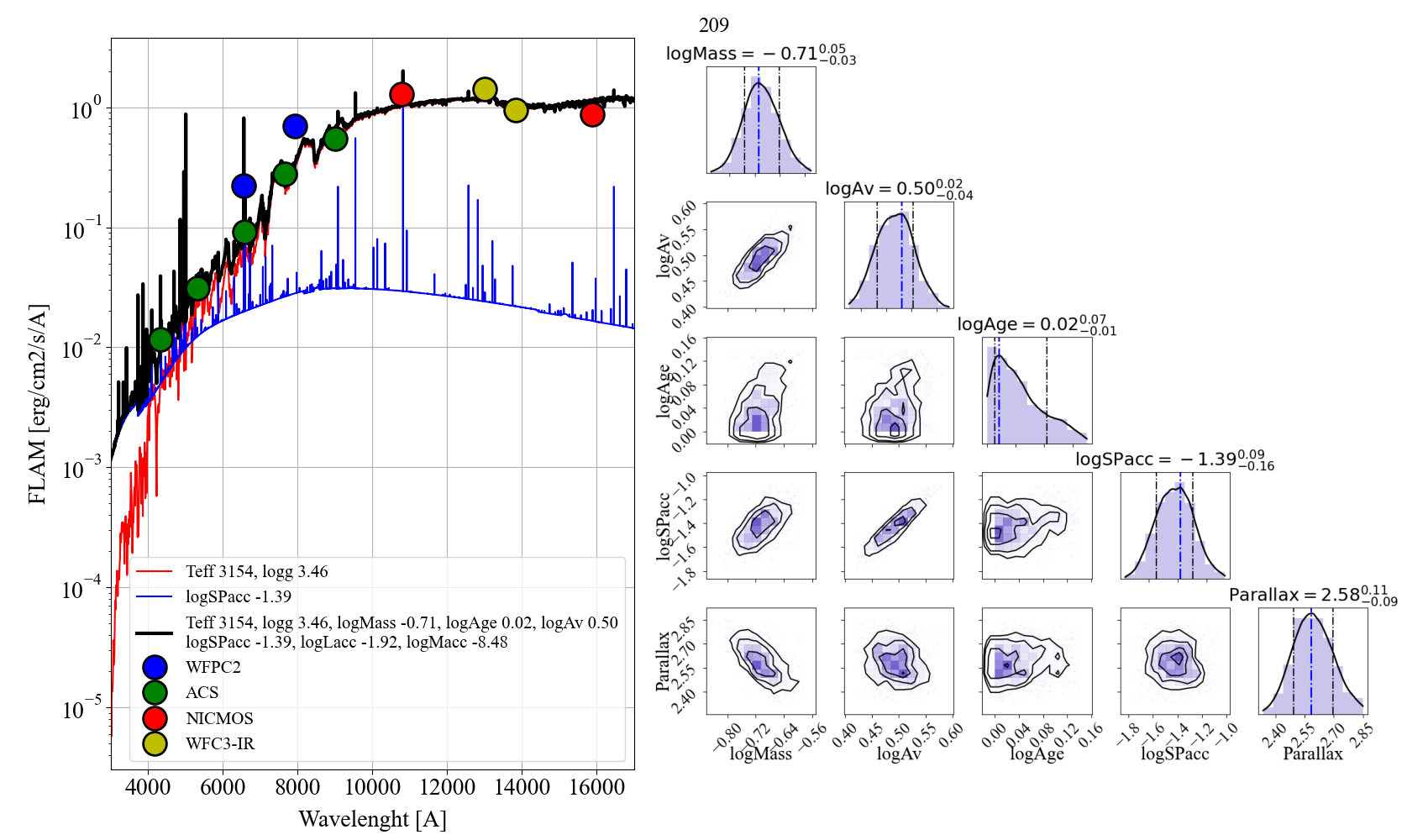 }
\caption{Sample of SED fitting (left) and corner plots (right) for cluster target sources in the catalog. The red line in the left panel shows the original spectrum for the star, while the blue line shows the slab model adopted in this work scaled by the $log_{10}SP_{acc}$ parameter. The black line instead represents the spectrum of the star combined with the slab model. The filters utilized for each fit are shown as circles color-coded by their respective instrument.
On the right panel, the peak (blue dotted line) and the limits of the $68\%$ credible interval (black dotted lines) are reported for each parameter. The black line in the histograms shows the Gaussian KDE.
The complete figure set (1049 images) is available in the online journal.}
\end{figure*}

\begin{figure*}[!t]
   \centering
   \includegraphics[width=0.99\textwidth]{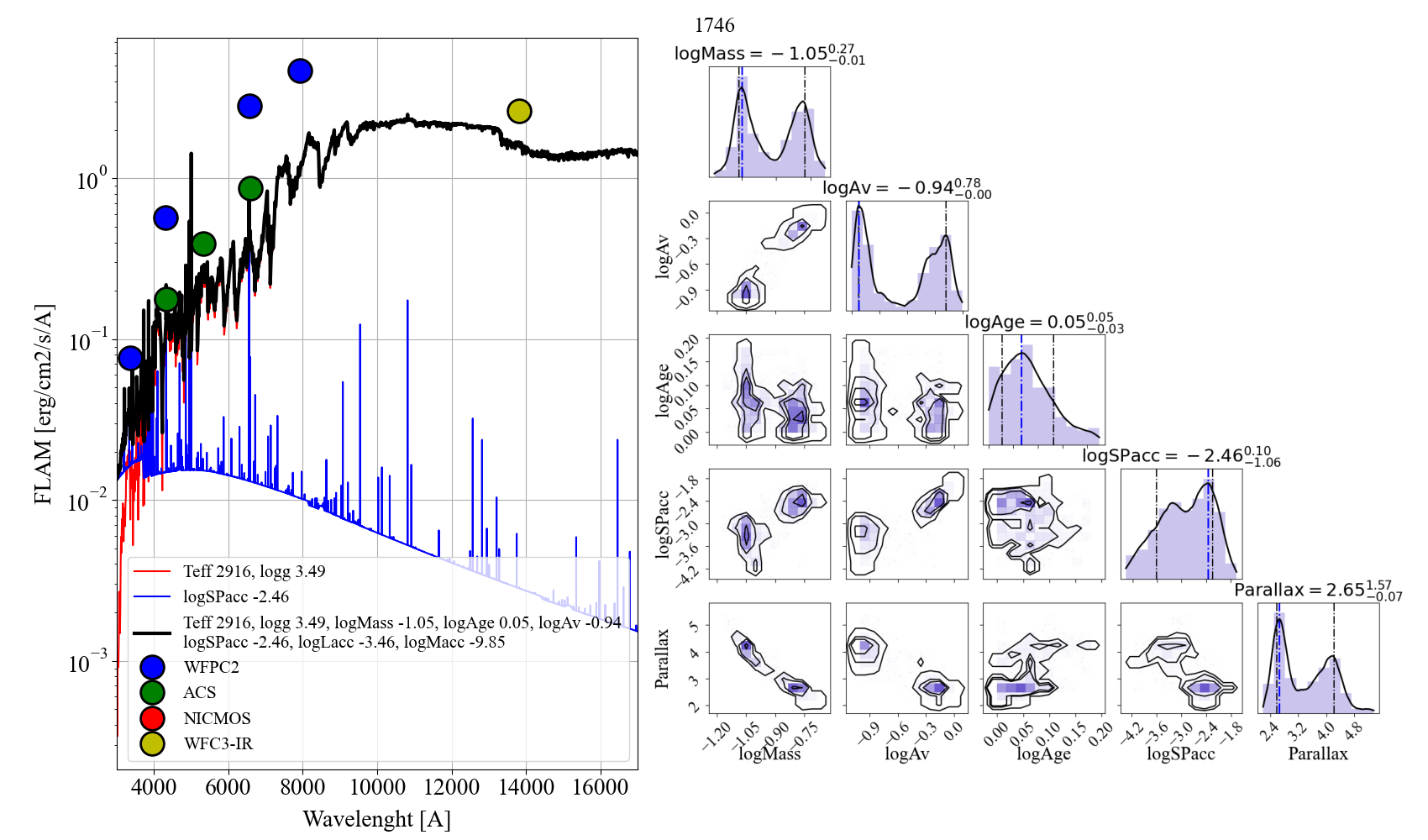}\\
   \includegraphics[width=0.99\textwidth]{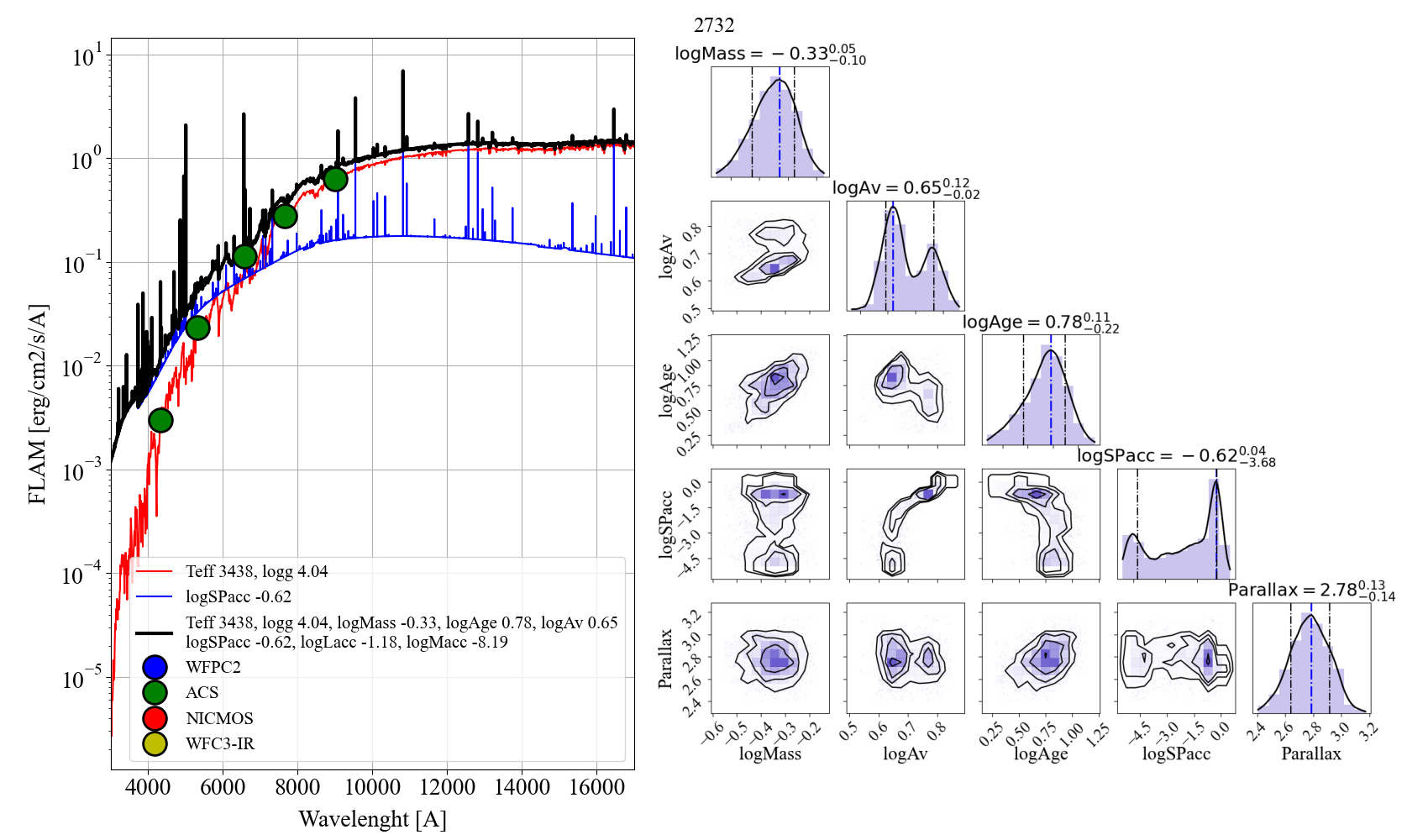}\\
   
\caption{Similar to Figure \ref{fig:corners}, this Figure show two examples of problematic outcome from our Bayesian fit. The top row shows a source detected in multiple filters, whose solution appears bimodal for almost all  parameters and that overall fails to match the observations. The bottom row instead shows the case of a source detected in only the ACS filters that appears to be bimodal in the Av and the $SP_{acc}$ parameters, and that produced an overestimated spectrum of accretion (blue line) that can not be accounted for by the other filters.}
   \label{fig:corners1}
\end{figure*}

Figures \ref{fig:corners} show examples of the SED fitting (left) and corner plots (right) representing the posterior probabilities for the parameters of two random stars. The red line in the left panel shows the original spectrum evaluated for the star, while the blue line shows the slab model adopted in this work scaled by the $log_{10}SP_{acc}$ parameter.The black line instead represents the spectrum of the star corrected by the slab model. The filters adopted for each fit are shown as circles color-coded by instrument. On average, our tests show that we are able recover the spectral type determined by \cite{Hillenbrand1997,Hillenbrand2013} within 2 subclasses (i.e., $\pm 200$~K).

Figure \ref{fig:corners1} instead, show examples of problematic fitting, generally due to the presence of a bimodal solution in one or more parameter posterior distributions, too few filters available to properly pinpoint the right spectrum to the photometry, the presence of an excess of flux in the filters most sensitive to accretion (e.g., F336W, F656N, F658N) not reflected in the other filters, or any combination of the above. This generally produces an over or under-estimate of the spectrum for the source.

Each corner plot shows the median value, which we adopted as the best-fit estimator, as well as the $68\%$ credible intervals of each fitted parameter's marginal posterior distribution. We define this interval so that the $68\%$ credible interval contains $68\%$ of the total probability, with (100 - 68)/2 = $16\%$ of the remaining probability on either side. This definition coincides with the usual $1\sigma$ interval for a Gaussian distribution. Note that the posterior distributions of our  parameters are generally not Gaussian nor symmetric in most cases. The marginal distributions for the single parameters are still useful for defining the parameter credible intervals but do not capture the whole information available in the full posterior distributions, with their correlation. 
For detailed studies of individual sources, a complete gallery is available in the online article.
\begin{figure*}[!ht]
    \centering
    \includegraphics[width=1\textwidth]{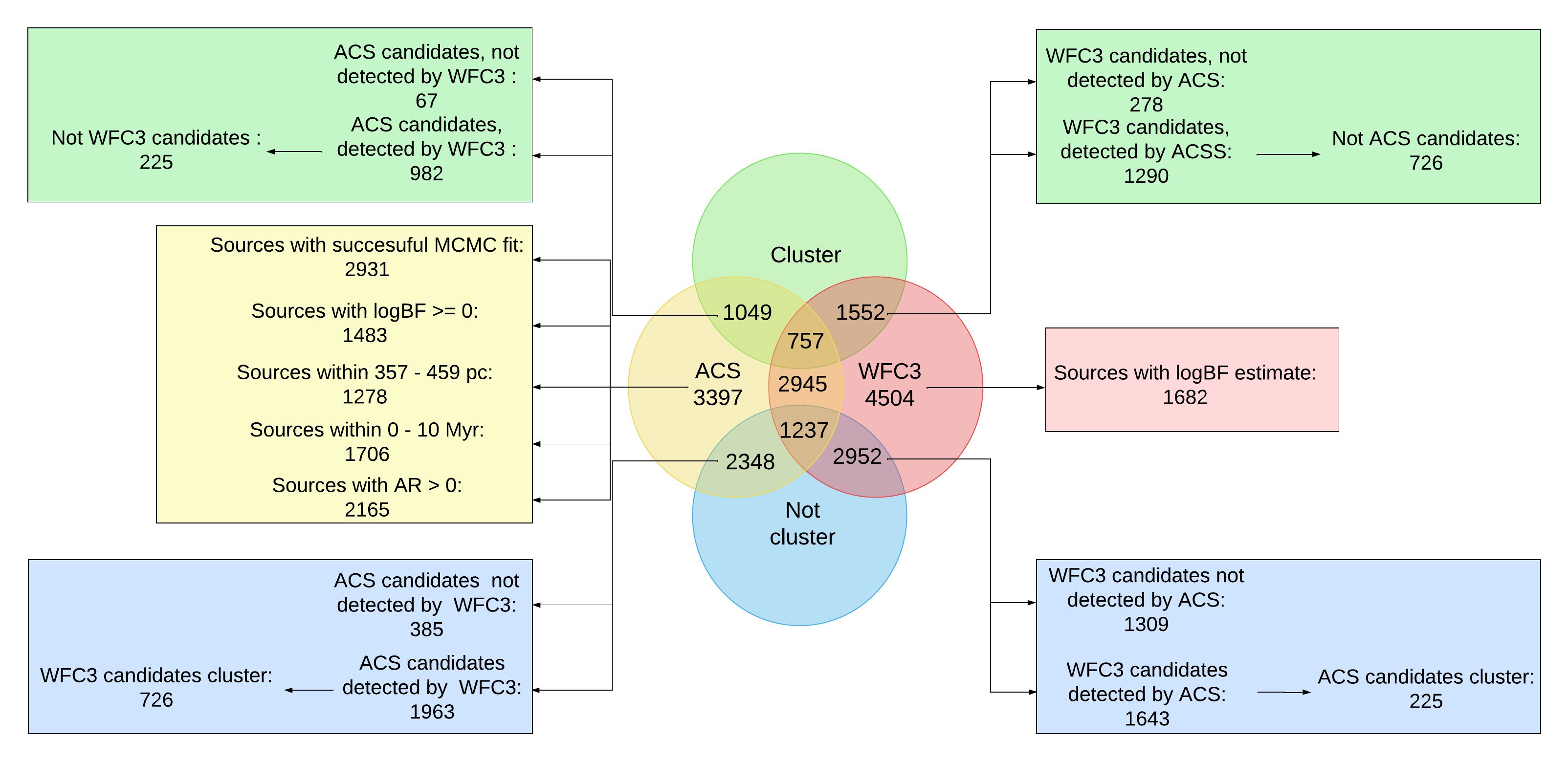}
    \caption{ACS/WFC3 ONC Venn diagram showing the overlap between ACS and WFC3 surveys and their respective characterization as candidate cluster/not cluster members.} 
    \label{fig:Venn_diagram}
\end{figure*}

\begin{figure*}[!ht]
    \centering
    \includegraphics[width=1\textwidth]{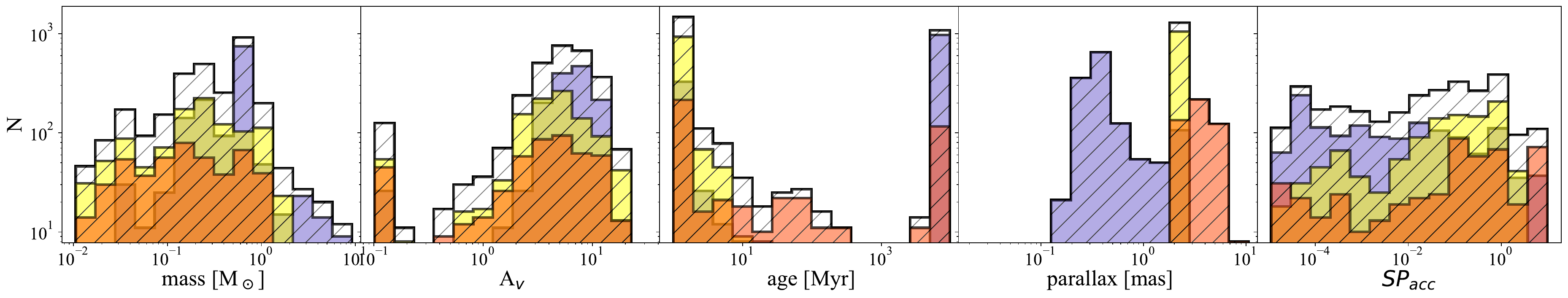}
    \caption{Histogram distributions for the five parameters (i.e.; mass, extinction, age, and parallax and $SP_{acc}$) obtained through the Bayesian analysis described in Section \ref{sec:Bayesian analysis} for all the sources in the catalog (white), cluster (\textit{reference catalog} - yellow), background (blue), and foreground (red) members. Note: larger parallaxes mean closer sources to the sun.} 
    \label{fig:Hists_all}
\end{figure*}

\section{Results}
\label{sec:Results}
\subsection{Membership selection}
\label{sec:Membership selection}
We will discuss in the following our approach to cluster membership selection.
On the basis of our analysis, we determine ONC membership as follows. 
First we eliminate all the sources where the MCMC fit couldn't converge, removing 468 sources from our sample and leaving us with a total of 2931 candidates.
Then, to be an ONC member:
\begin{itemize}
    \item The derived parallax of a candidate member must be compatible with $2.5\pm 0.3$ mas ( $402^{+57}_{-45}$ pc),  i.e. $1\sigma$ from the peak of the Gaia DR3 parallax distribution for all the sources falling in our field of view. This leave us with 1278 candidate sources from the initial 2931.
    Taking full advantage of the posterior distribution of the parallax solution (PDPS) of each source, we evaluate the ratio of the area under the curve (AOC) of PDPS that lies within the same $2.5\pm 0.3$ mas  interval over the total AOC of the PDDS of each source. This ratio (that we will call \textit{area ratio} or AR) provides us with an estimate of the probability that the two parallaxes are compatible. 
    \item The derived age of a candidate member must be compatible with a recent star formation event, i.e.  $\leq 10$ Myr. This captures the age range of interest, as 
    previous works  \citep[e.g.,][]{Jeffries2011,DaRio2012, Beccari2017} show that the typical age for this cluster is $\sim 2$ Myr, with an age spread at most of a few Myrs.
    In particular, we find that $\sim 75 \%$ of our candidates have an age $\lesssim 2$ Myr, while if we cut the age distribution at $\lesssim 10$ Myr, $\sim 88\%$ of the selected sources are younger than $\lesssim 2$ Myr and $\sim 97\%$ are younger than $\lesssim 5$ Myr. The age selection leads to discarding other 199 sources from our candidate sources catalog, leaving us with a candidate cluster catalog of 1079 sources. 
    \item After visually inspecting each SED fit of these 1079 cluster members, we exclude 30 candidates 
    due the gross inconsistency between the observed photometry and the best fit (see examples in Fig. \ref{fig:corners1}), reducing the number of candidates cluster member to 1049.
\end{itemize}
This sample of 1049 sources distilled from the initial 3399 stars, i.e. $\sim 31\%$ of the  initial sample, will represent our reference catalog of candidate cluster members (or reference catalog for short). Their AR distribution shows a peak at a median value AR$\sim 0.89^{+0.11}_{-0.17}$. This confirm that the majority of members in our reference catalog have high probability that their estimated distance distribution is compatible with the Gaia DR3 distance distribution for the ONC of $402^{+57}_{-45}$ pc.
To assess the number of stars that may have been incorrectly classified as background objects, we generated a synthetic photometry catalog of background stars using Synphot. Starting from the stellar parameters T$_{eff}$, $\log Z$, $\log g$ and distance provided by a Besançon model of the Milky Way at the coordinates of the ONC, we determined the photometry of each source using \textit{phoenix} spectra. For each filter, the magnitudes were extracted after reddening the spectrum using the sum of the extinction provided by the Besançon model and the Scandariato's extinction map for the OMC \citep{Scandariato2011}, for a randomly generated position within our field of view. 
Comparing the number of background stars with the Besançon predictions,  we are able to account for $\sim 90\%$ of the predicted background stars in the ACS redder filters and $\sim 80\%$ in the ACS blue and WFC-IR filters filters. Overall, this test suggest that the significant fraction of background sources determined by our approach is compatible with the predictions from the Besançon model, within the uncertainties on the extinction due to the non-uniform column density of the OMC.

As anticipated in Sec. \ref{sec:Observation and Data Reduction}, we cross-match the ACS (yellow) and 
WFC3 (red) surveys. Discussing the WFC3 dataset, \cite{Robberto2020} present a detailed Bayesian analysis to disentangle bona-fide low-mass stars and substellar cluster members from background sources based on the presence of water in absorption shown by F130N-F139M color.

Figure \ref{fig:Venn_diagram} shows the overlap between ACS and WFC3 surveys and their respective characterization as candidate cluster (green) and not cluster members (blue - either background, foreground or where the sorting approach failed). It turns out that of 4504 WFC3 sources, only 2945 are also present in our ACS catalog (out of a total of 3399 ACS initial sources) or, in other words, $\sim 87\%$ of ACS sources overlap with WFC3.
We must remind the reader that in the region of overlap between ACS and WFC3 some sources can be labeled cluster in one survey and not cluster in the other, and vice-versa. Indeed, the total number of overlapping sources (2945) is not equal to number of overlapping candidates cluster (757) plus the number of the overlapping not candidates cluster (1237), as we need to include also the ACS candidates labeled as not cluster by WFC3 (225) and the WFC3 candidates labeled as not cluster by ACS (726).

\begin{figure*}[!th]
    \centering
    \includegraphics[width=1\textwidth]{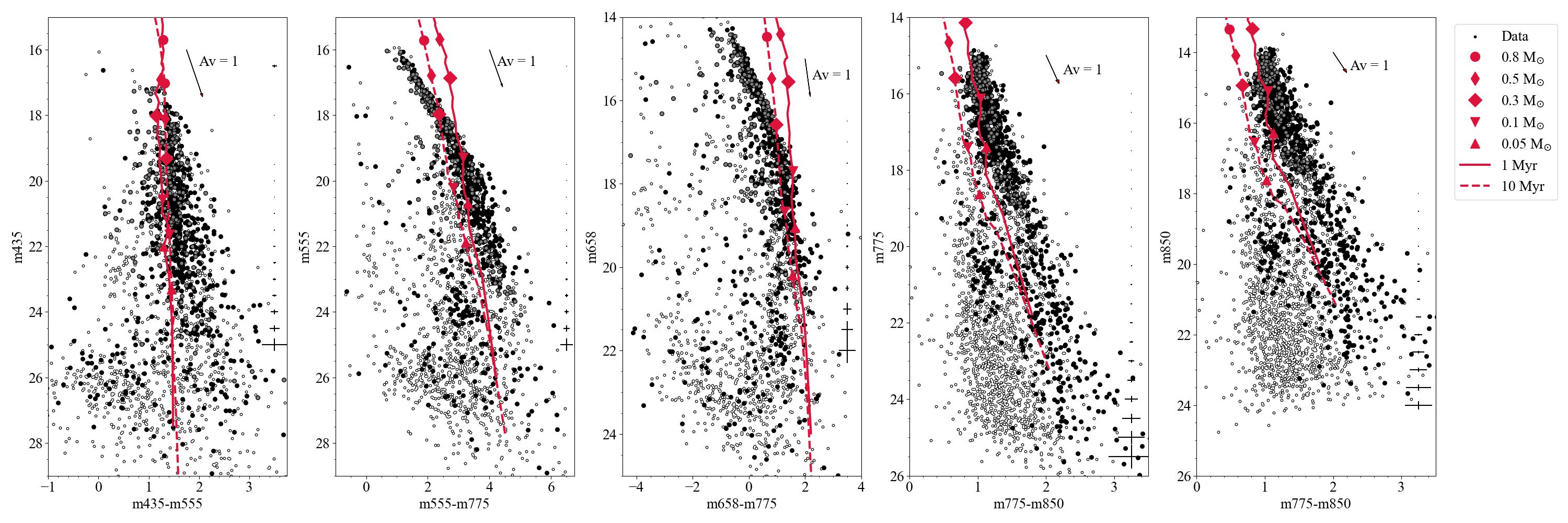}\\
    \includegraphics[width=1\textwidth]{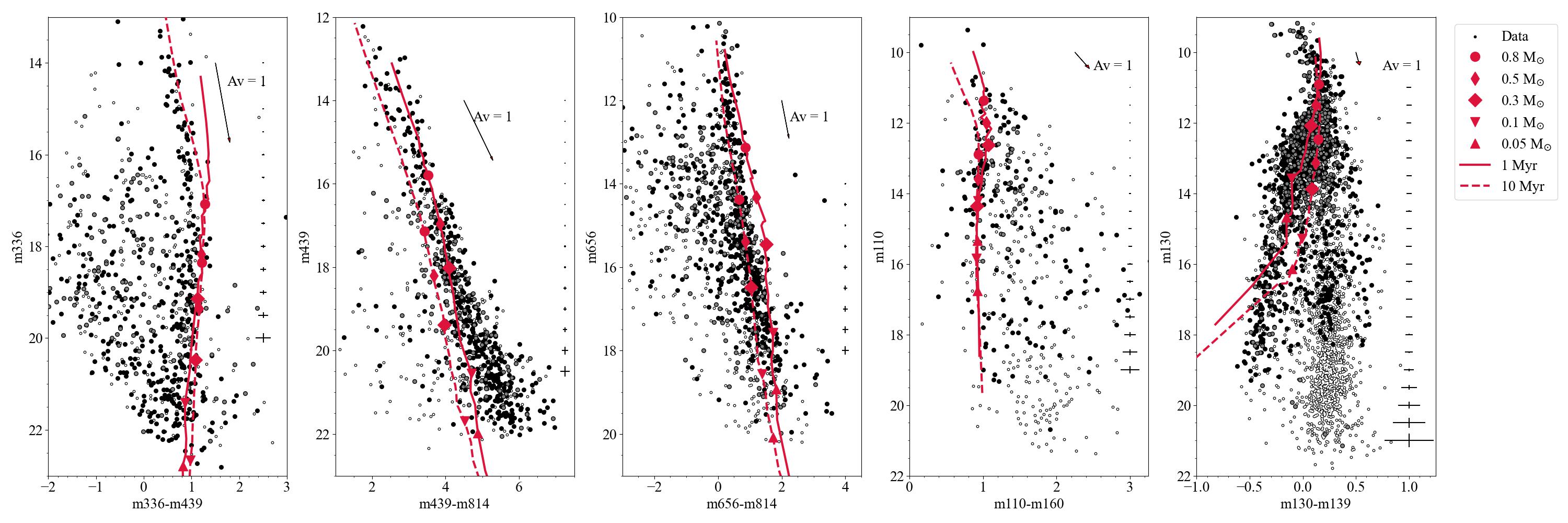}\\
    \caption{CMDs for ACS, WFPC2, NICMOS, and WFC3-IR filters for the whole sample of unique sources in the catalog. The dark dots mark the reference catalog sources, while the grey ones mark the the new sources added to create the expanded catalog.
    The red continuous and dashed-dotted line shows the 1 and 10 Myr isochrone for each CMD. Typical error bars averaged for bins of magnitude are shown on the right of each plot. 
    \label{fig:CMD_cluster}}
\end{figure*}

From the analysis of the ACS reference catalog we find 67 sources not listed in the WFC3 catalog, while 225 sources are instead discarded by \cite{Robberto2020} on the basis of their position on the CMD. Those are recovered by our MCMC analysis thanks to a compatible solution with the Gaia DR3 estimated distance and age of the cluster.

Analyzing the WFC3 sample, out of 4504 sources we identify 1552 candidate cluster members on the basis of the Bayes Factor (BF) $\log(BF)>= 0$, corresponding to a $\sim 34\%$ ratio very similar to the $\sim 31\%$ we reported for the ACS survey.
Of these 1552 sources, 278 are not listed in our initial ACS catalog. Of the remaining 1290 WFC3 candidate sources, 726 are not identified as ACS candidate sources by our method, either because the MCMC fit puts them out of the cluster (418 cases) or because it fails to converge to a solution.
 
Since we cannot exclude that both the 308 sources missing the MCMC fit and the 278 sources not detected by our survey \citep[but still identified as candidate cluster by ][]{Robberto2020} are actually part of the cluster, we will add them back creating a second \textit{expanded catalog} of candidate cluster members.
Lacking the full distributions provided by the MCMC analysis, we will adopt the values for the mass, extinction and photometry from \cite{Robberto2013} for these extra sources, but since only 497 of the 586 missing sources have an estimate of the mass and extinction, we will add only these last sample to our expanded reference catalog, counting 1546 sources in total.

\begin{deluxetable*}{ccccccccccccc}
\tabletypesize{\footnotesize}
\setlength{\tabcolsep}{3pt}
\tablecaption{\textit{Sample of the physical parameters of all the sources included in this work.}
\label{Tab:char}}
\tablehead{
\colhead{id} & \colhead{membership} &
\colhead{Ra} & \colhead{Dec}  & \colhead{mass} & \colhead{\Av} &
\colhead{Age}  & \colhead{Dist} & \colhead{T} &
\colhead{logL}  & \colhead{R}  & \colhead{logg}\\
\colhead{}   & \colhead{} &
\colhead{(deg)}   & \colhead{(deg)}  & \colhead{(\Msun)} & \colhead{(mag)}  & \colhead{(Myr)}   &  \colhead{(pc)} & \colhead{($^\circ$K)}  &  
\colhead{(\Lsun)} &  \colhead{(Gcm)}   &  \colhead{(cm/s$^2$)}  
}
\startdata
1   & n   & 83.541204 &-5.373961 &
            $0.6^{+0.1}_{-0.3}$     & 
            $4.2^{+0.7}_{-1.6}$     &
            $7014.6^{+1197.5}_{-2224.4}$&
            $2995.3^{+1454.2}_{-1473.4}$&
            $3808.3^{+501.5}_{-446.2}$&
            $-1.3^{+0.6 }_{-0.1}$   &
            $36.8^{+11.7}_{-19.8}$  &
            $4.7^{0.2}_{0.2}$ \\
8   & y   & 83.550104 &-5.405411    &
            $0.4^{+0.1 }_{-0.1}$    &
            $3.0^{+0.3}_{-0.6}$     & 
            $5.1^{+1.2}_{-1.3}$     & 
            $361.0^{+12.5}_{-12.6}$ & 
            $3387.8^{+76.5}_{-87.6}$& 
            $-0.9^{+0.3}_{-1.0}$    & 
            $72.2^{+0.5}_{-1.8}$    & 
            $4.0^{+0.3}_{-0.7}$     \\
\enddata
\tablecomments{\textit{Table \ref{Tab:char} is published in its entirety in the machine-readable format. A portion is shown here for guidance regarding its form and content. Note that in the published machine-readable version of this table, the errors will occupy a different column.}}
\end{deluxetable*}


Figure \ref{fig:Hists_all} shows the histograms of the estimated values of each parameter for all the sources in our catalog. The mass, age, and distance panels, in particular, show a mix of low-mass ($\lesssim 0.3$ \Msun) and more massive ($\sim 1$ \Msun) stars, of very young ($\lesssim 10$ Myrs) and very old ($\gtrsim$ 1 Gyr) stars, as well as close ones (log(parallax) $\sim 0.4$ mas corresponds to $\sim 400$ pc) and far away (log(parallax) $>> 0.4$ mas). All these distributions are compatible with two distinct populations: one younger and closer to us, the ONC (yellow), and one older and far away (the background stars - blue).
Moreover, the parallax distribution shows a small population of sources with a parallax much smaller than the one expected for ONC cluster member (red - log(parallax) $> 0.4$ mas) compatible with the presence of an additional population of foreground objects that are indeed expected on the line of sight.

For the candidates cluster distributions from the \textit{reference catalog}, the median values of each distributions are: mass $0.21$ \Msun; $A_V$ $4.35$ mag, age $1.07 $ Myr, parallax $2.52 $ mas, and SP$_{acc}$ $0.08 $.

Figure \ref{fig:CMD_cluster} shows the color  magnitude-diagrams  (CMD) for eight combinations of simultaneously observed bandpasses, i.e.
ACS: m435\,vs.\,m435-m555, m555\,vs.\,m555-m775,  m775\,vs.\,m775-m850, 
and m850\,vs.\,m775-m850, on the top row and 
WFPC2: m336\,vs.\,m336-m439, m439\,vs.\,m439-m814, m656\,vs.\,m656-m814, 
NICMOS: m110\,vs.\,m110-m130, 
and WFC3-IR: m130\,vs.\,m130-m139 
on the bottom row. All sources with photometry in each given pair of filters are plotted, together  with  our model 1 Myr and 10 Myr isochrones and the $A_V= 1$ reddening vector, for comparison.  Cluster  members from the reference catalog are plotted as filled black circles, while the additional sources from the expanded catalog are marked in grey. Non-members (either because they can be tagged as field stars or due to unreliable estimates of the source's properties) are open circles. Grey circles mark instead the new sources added to create the expanded catalog as discussed above.
In all these diagrams the positions of the cluster members selected using the criteria explained above are generally compatible with the locus of young stellar objects (once extinction and accretion are taken into account). The systematic departures at the bright end are due to the different saturation thresholds in the various filters.

From  the masses and ages derived by the fitting routine, we have also retrieved the corresponding stellar luminosity (\Lstar) and effective temperature (\Teff) interpolating over our family of BT-Settl  isochrones.  The physical parameters estimated through our Bayesian analysis for all 3399 sources of our main catalog are provided in Table \ref{Tab:char}. 
The breakdown of the table is as follows: the first two columns show the ID as provided from the \textit{Stra}KLIP pipeline for cross-identification and the membership flag. The following columns show R.A. and DEC, followed by the estimated parameters Mass, \Av, Age, distance of the source from the Sun, and derived parameters \Teff, log\Lstar, $R$ and $logg$ with relative upper and lower error limits. 

\begin{deluxetable}{ccc}
\tabletypesize{\footnotesize}
\setlength{\tabcolsep}{15pt}
\tablewidth{10pt}
\tablecaption{\textit{Sample of the accretion properties of all the sources included in this work.}
\label{Tab:acc}}
\tablehead{
\colhead{id} & 
\colhead{logL$_{acc}$} & \colhead{log\dMacc} \\
\colhead{} &
\colhead{(\Lsun)} & \colhead{(\Msun yr$^{-1}$)}
}
\startdata
1        & $-5.051^{+4.27}_{-0.054}$ &  $-12.468^{+4.319}_{-0.059}$ \\
8        & $-2.178^{+0.43}_{-0.022}$ &  $-9.137^{+0.388}_{-0.029}$ \\
... \\
\enddata
\tablecomments{\textit{Table \ref{Tab:acc} is published in its entirety in the machine-readable format. A portion is shown here for guidance regarding its form and content. Note that in the published machine-readable version of this table, the errors will occupy a different column.}}
\end{deluxetable}

\subsection{Accretion luminosity and mass accretion rate}
\label{sec:Accretion luminosity and mass accretion rate}

In this Section we will discuss the process adopted to evaluate \Lacc and \dMacc.

 
\begin{figure*}[!ht]
    \centering
    \includegraphics[width=0.99\textwidth]{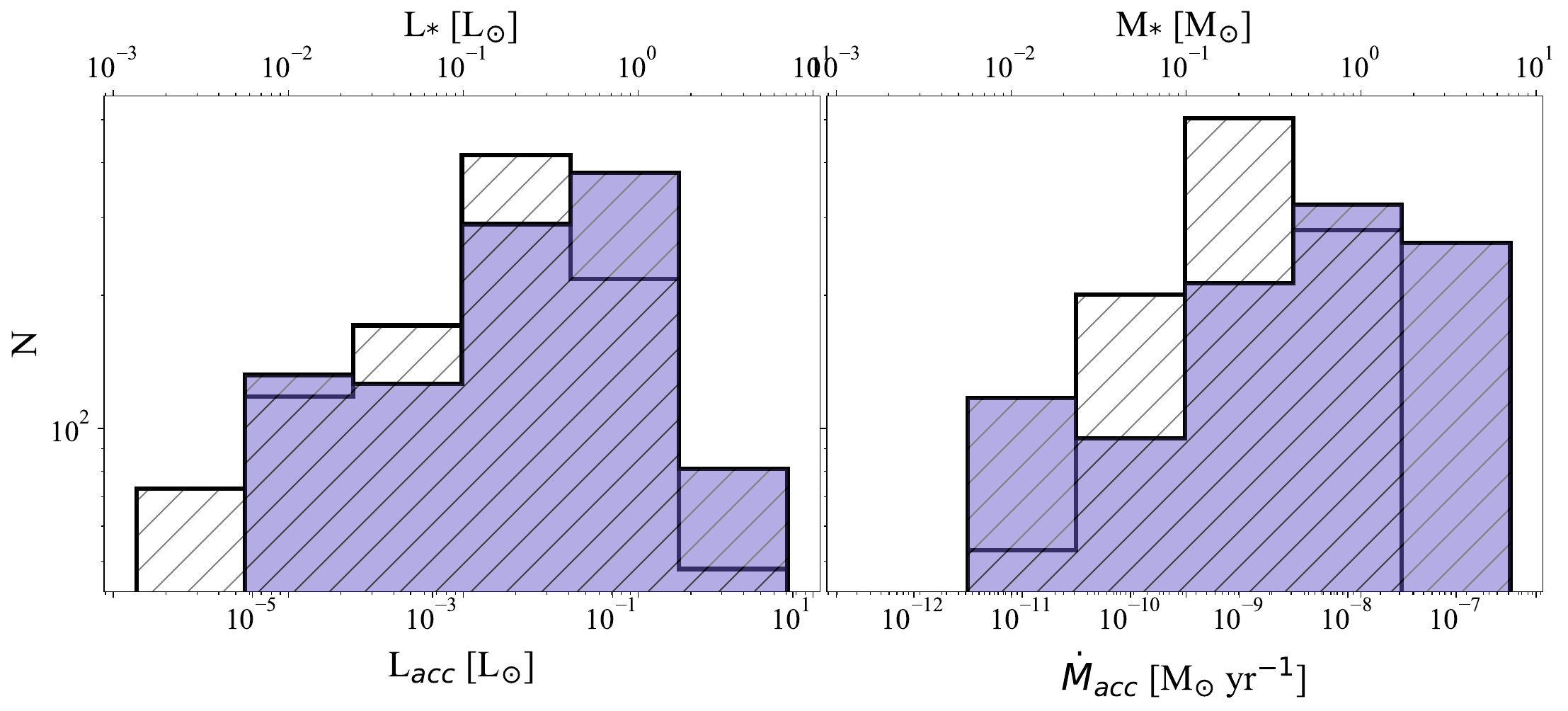}
    \caption{Histograms of the accretion luminosity \Lacc (right) and mass accretion rate \dMacc (left) derived by the stellar parameters obtained from our Bayesian analysis (blue) with \Lstar and \Mstar overplotted on a twin axis respectively (white).}
    \label{fig:acc_hist}
\end{figure*}

As mentioned earlier in Section \ref{sec:Bayesian analysis}, to model the intensity of the accretion spectrum in our Bayesian SED fitting we introduce an accretion parameter, $SP_{acc}$, that represents the fraction of stellar bolometric luminosity that can be attributed to accretion. Having derived \Lstar, one immediately obtains the accretion luminosity (\Lacc) and therefore the mass accretion rate (\dMacc) by inverting the free-fall equation that links the luminosity released in when the accretion flow impacts the stellar surface with the rate of mass accretion:
\begin{equation}
    \Lacc \simeq\frac{G\Mstar\dMacc}{\Rstar}\bigg(1-\frac{\Rstar}{R_{in}}\bigg)
    \label{eq:dMacc}
\end{equation}
where \Rstar and R$_{in}=5$ R$_{\star}$ are the star and inner-disc radius \citep{Gullbring1998,Hartmann1998}. 


In Table \ref{Tab:acc} we list the accretion properties determined for our catalog. The breakdown of this table is as follows: the first column shows the ID as provided from the \textit{Stra}KLIP pipeline for cross-identification. The next columns list the derived accretion parameters log\Lacc and log\dMacc and relative errors for each of our sources. 

Focusing on the cluster members, in Figure \ref{fig:acc_hist} we show the distributions of \Lacc and \dMacc. In solar units, the median values are: \logLacc $-1.62 \pm 1.43$ and \logdMacc $-8.29 \pm 1.39$, respectively.
On the same plot, using a twin axis as a reference, we also show the corresponding \Lstar and \Mstar distribution. 

Our estimates of \Lacc and \dMacc are probably affected by some selection bias inherent to the nature of the dataset we are analyzing. While it is generally true that the number of filters plays a major role in the goodness of the fit, this is particularly true for the estimate of the accretion luminosity, in particular, if the missing filters are those most sensitive to the specific characteristics of the accretion spectrum, i.e. F336W (an HST equivalent to the Johnson U band) and the F656N and F658N filters centered on the $\Ha$~line. Photometry in those bands may be missing either because the source is too faint, or even too bright with strong accretion even leading to saturation. In this last case, the other bandpasses  may provide an indication of accretion, but inaccurate or underestimated values. Also, and especially in the inner regions of the ONC, the presence of circumstellar emission (e.g. proplyds) can influence the photometry in these same filters, leading to an overestimate of the accretion luminosity. 
Indeed, even though we removed from this analysis the known, spatially resolved proplyds, we cannot completely exclude the presence of circumstellar emission in unresolved stars.

\section{Discussion}
\label{sec:Discussion}

\subsection{Three-dimensional spatial distribution}
\label{sec: Three-dimensional spatial distribution}

\begin{figure*}[!ht]
    \centering
   \includegraphics[width=0.45\textwidth]{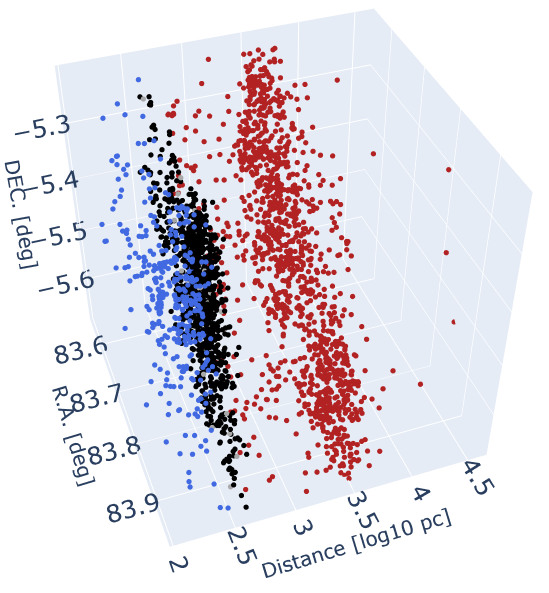}
    \includegraphics[width=0.54\textwidth]{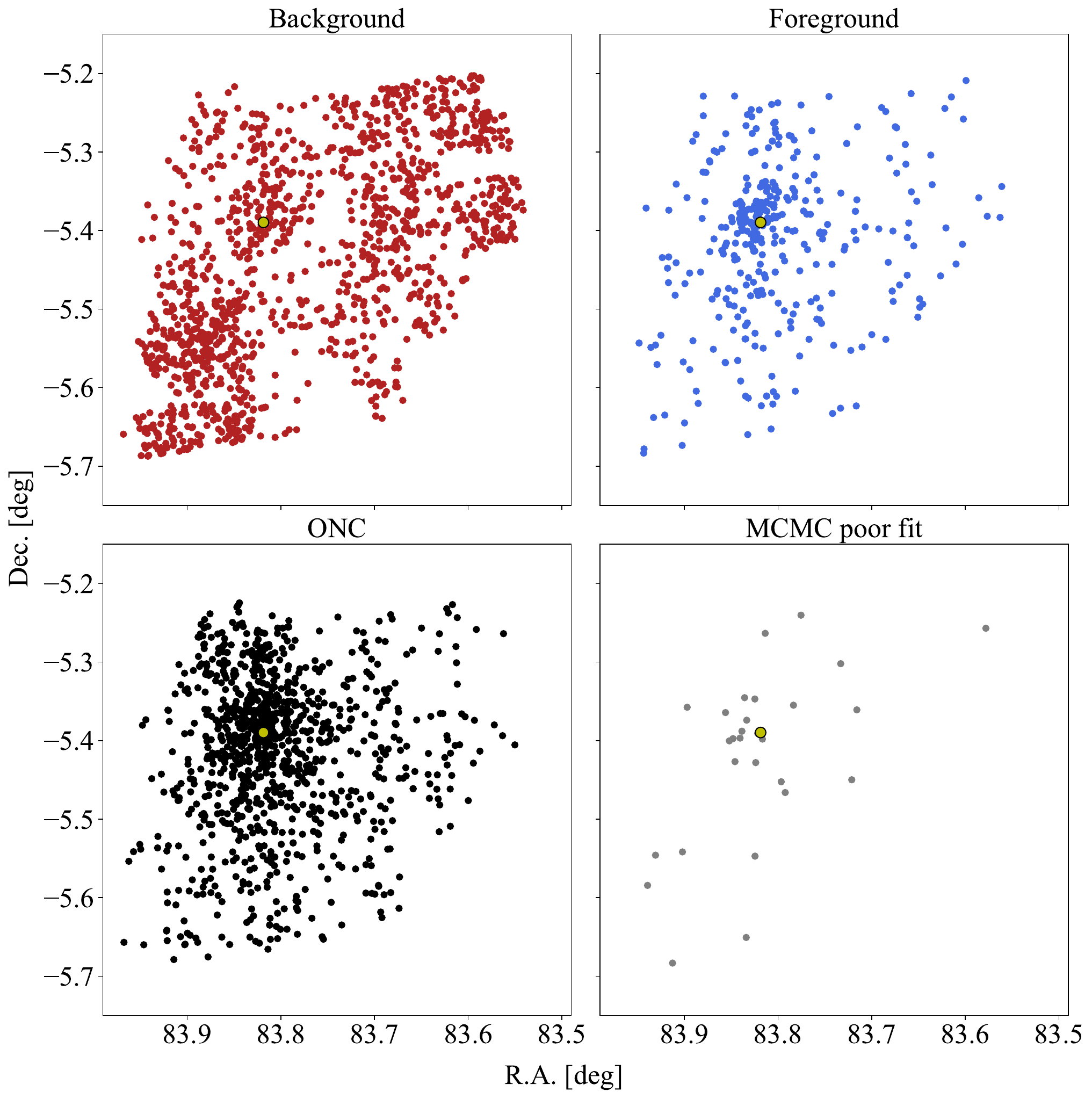} 
    \caption{Spatial distribution of sources in the catalog divided by membership. Left: 3-d spatial representation of foreground (blue), candidate ONC members (black), candidate ONC members with not converging MCMC fit (gray) and background (red). An interactive 3-d version of this plot is provided in the online version of this paper. Right: 2-d view for the same groups. A yellow marker shows the position of $\theta^1$ Ori-C on each plot.
    }
    \label{fig:Spatial_dist}
\end{figure*}

The distances determined by our MCMC Bayesian routine allow reconstructing the three dimensional spatial distribution of the sources in our reference catalog. The left panel of Figure \ref{fig:Spatial_dist} shows the  three main populations that can be discerned: the candidates background sources (red dots - 1303 sources) to which the fitting procedure assigns an average  distance, $d\sim 3$~Kpc  ($\log d$(pc)$\sim 3.6$), the candidates cluster members at about 400~pc ($\log d$(pc)$\sim 2.6$, black dots - 1044 sources), and the population of candidates foreground sources (blue dots - 350 sources) at $d\sim 250$~pc  ($\log d$(pc)$\sim 2.4$). We include in the cluster sample the 30 sources with bimodal MCMC solutions discussed in Section \ref{sec:Membership selection}, for which we have adopted the distances given by the highest maxima.
The right panel of the same figure shows the spatial maps of the 3 populations, together with the sources with bimodal solutions. For the candidate cluster population the 2-d spatial distribution shows a remarkable concentration toward the center (identified with $\theta^1$ Ori-C). The foreground population appear to be more uniformly spread {\bf , with a residual clustering toward the center of the cluster.
One may notice that a higher concentration of foreground sources toward the nebular core is also visible in the spatial map of the Gaia foreground stars. 
In fact, $77\%$ (189 sources out of 248) of our foreground sources match with Gaia foreground stars, indicating that the anomaly can be mostly driven by the Gaia prior, while $6\%$ (16) of them are assigned to the background by Gaia, and $17\%$ (43) are assigned to the cluster. For these 43 sources that Gaia identifies as clusters, $\sim 40\%$ of them still have a distance compatible with the cluster within the errors provided by our fit.}

The distribution of the background population shows instead a clear drop of source density along a N-S lane, a feature tracing the high-extinction caused by the main filament of the OMC1 molecular cloud \citep{Scandariato2011}.

\begin{figure*}[!ht]
    \centering
    \includegraphics[width=0.49\textwidth]{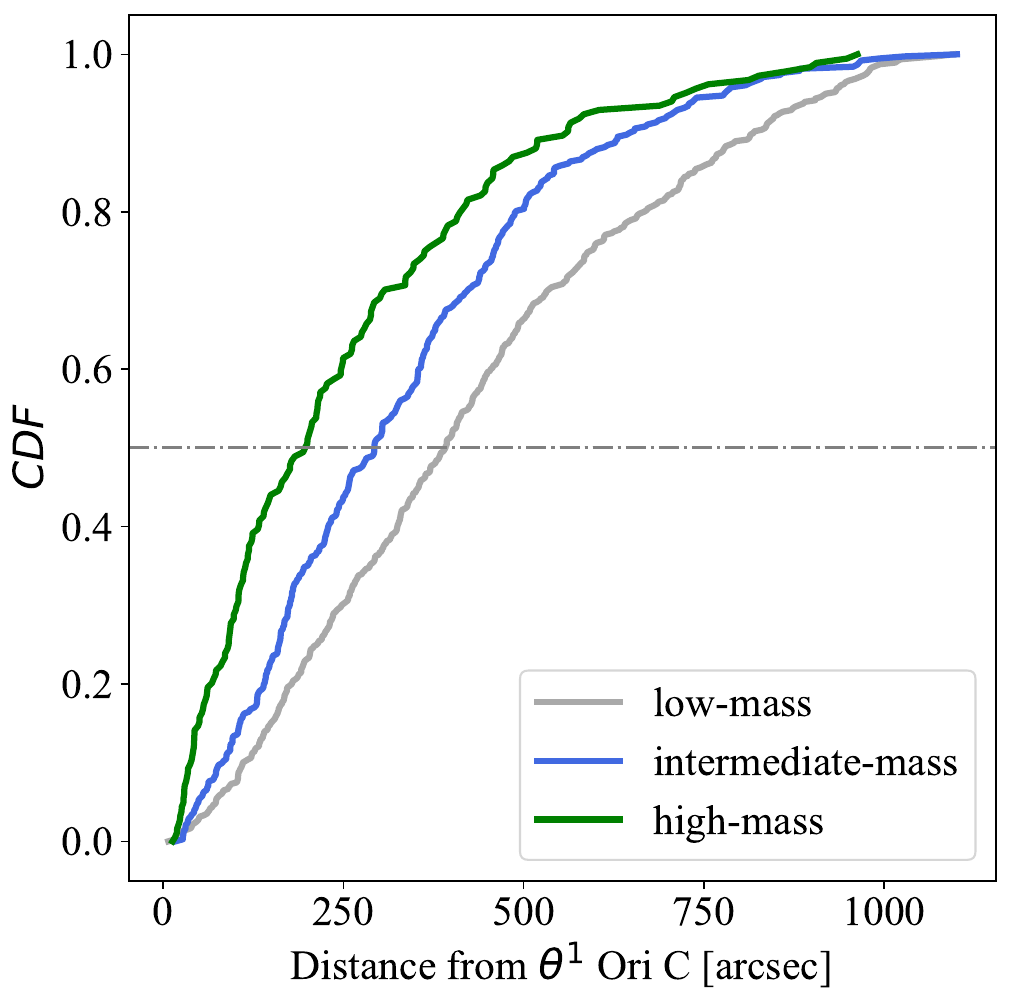}
    \includegraphics[width=0.49\textwidth]{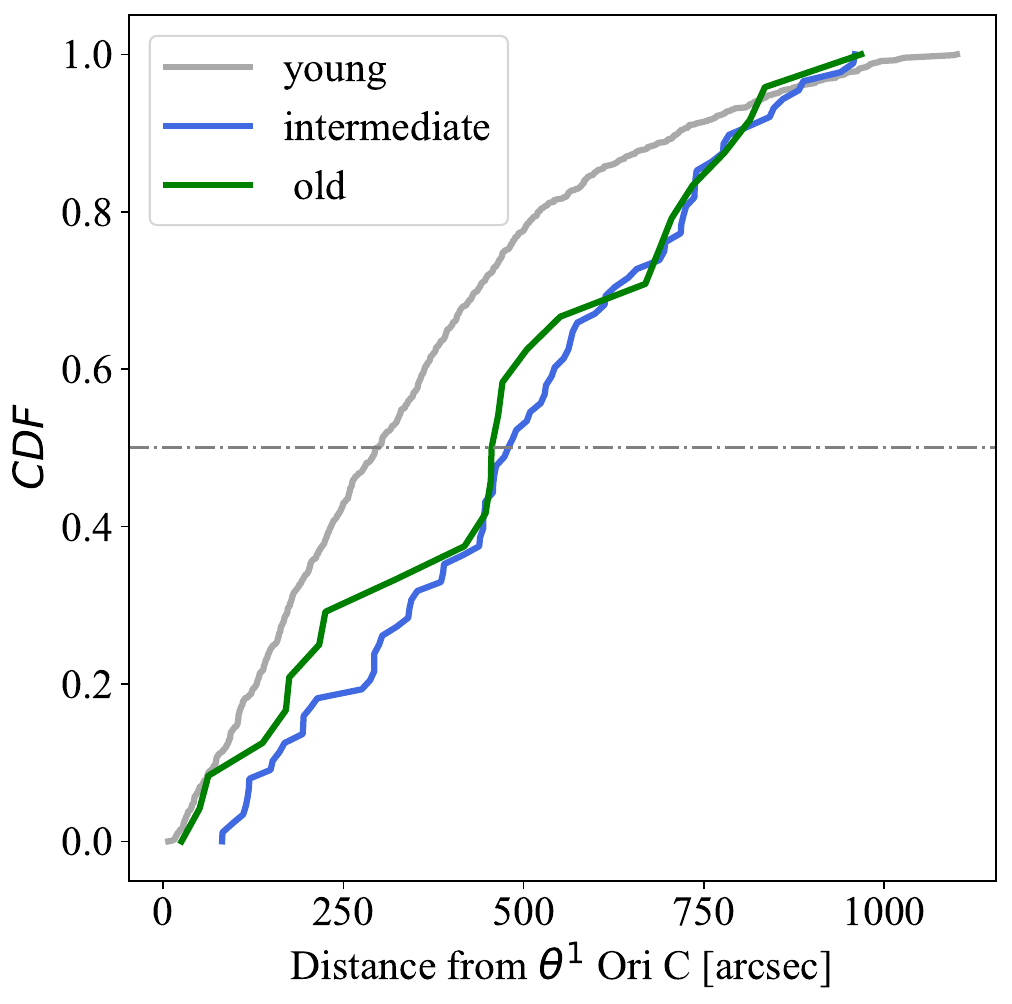}
    \caption{Cumulative distribution functions for sources in  our reference catalog of cluster members as a function of distance from $\theta^1 OriC$ (first three plots). Left: high-mass (mass $> 0.7$ \Msun - green), intermediate (mass between $0.2 - 0.7$ \Msun - blue) and low-mass (mass $< 0.2$ \Msun - gray) sources. 
    Right: old (age $> 5$ Myr - green), intermediate (age between $2 - 5$ Myr - blue), and young (age $< 2$ Myr - gray) sources.  
    }
    \label{fig:cdfs}
\end{figure*}

To assess how the main parameters of the low-mass objects probed by our survey vary with the radial distance from the center,  we present in Figure \ref{fig:cdfs} the cumulative distribution functions (CDFs) of the number of  candidate cluster members vs. distance, grouped according to their masses (left) and ages (right). 
The left panel in Figure \ref{fig:cdfs} indicates that the more massive ( $M>0.7~\Msun$, 185 sources, green line) have stronger tendency to cluster toward the central region than intermediate mass sources ($0.2<M<0.7~\Msun$, 383 sources, blue line), and these in turn are more clustered than the low-mass stars 
mass ($M<0.2~\Msun$, 481 sources, gray line), that tend to be the most dispersed. In particular, $50\%$ of the more massive sources are within about $200$~arcsec (0.4~pc) vs. $290$~arcsec (0.56~pc) for
the intermediate mass sources $390$~arcsec (0.75~pc) for the lowest mass objects. This trend indicates that mass segregation in the ONC extends to sub-solar masses, possibly down to 0.2~\Msun, with a degree of concentration decreasing with decreasing mass, extending into the sub-solar and possibly brown-dwarf mass range, similar to what found for masses $> 1-2~\Msun$ (not probed by our survey) by \cite{Hillenbrand1998}.

Similarly, the right panel of Figure \ref{fig:cdfs} presents the same CDF for three bins of ages.  While old ($> 5$ Myr - green - 25 sources) and intermediate age sources (between $2 - 5$ Myr - blue - 89 sources) appear similarly dispersed through the cluster, the large majority of young sources (age $< 2$ Myr - gray - 935 sources) show a higher degree of concentration toward the central region. This result sets a rather strong upper limit to the age of the cluster.

\subsection{Hertzsprung–Russell Diagram }
\label{sec:Hertzsprung–Russell Diagram }
\begin{figure}[!th]
    \centering
    \includegraphics[width=0.5\textwidth]{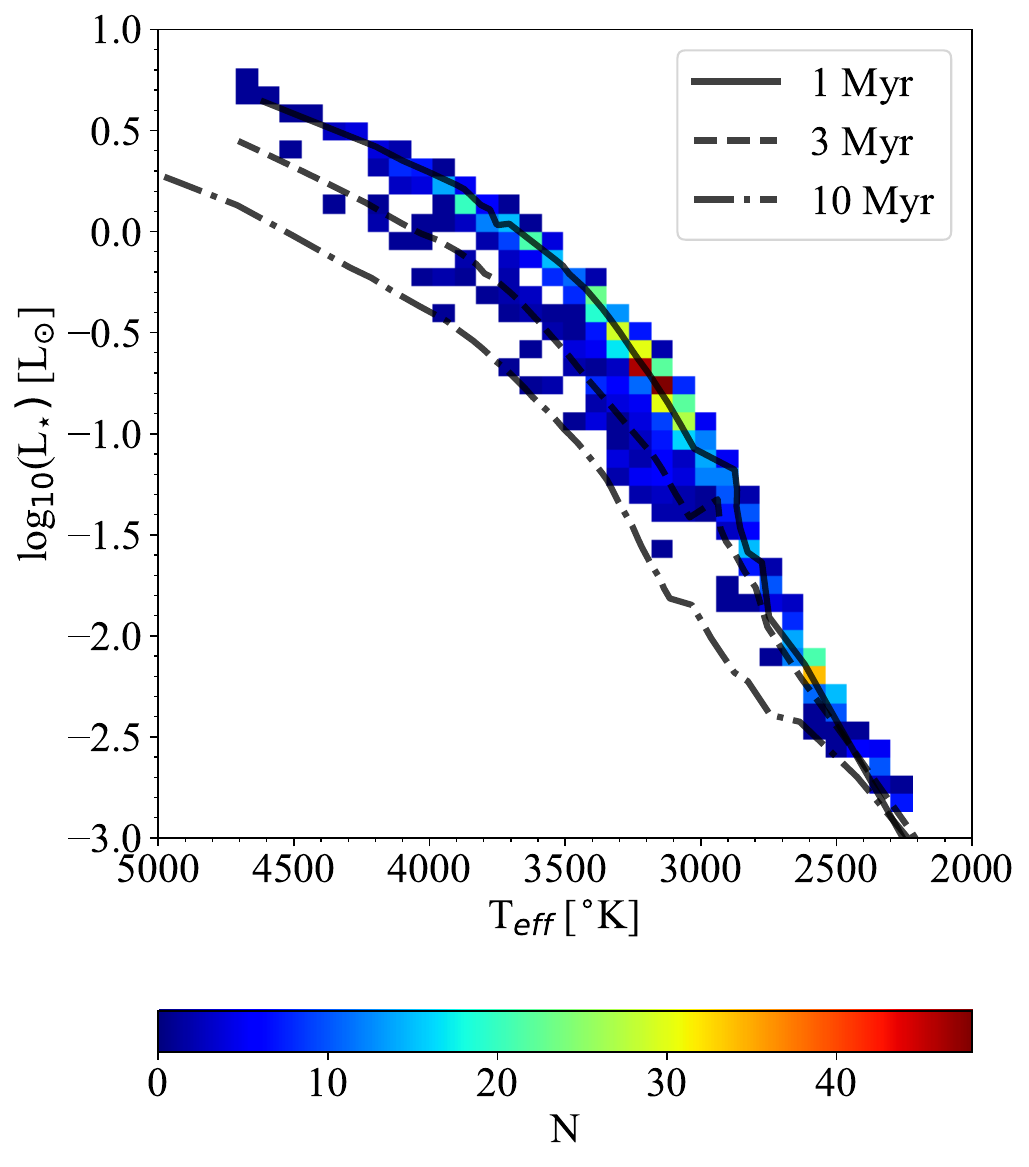}
    \caption{Hess diagram for the HRD for ONC candidate cluster sources. Luminosity and temperature are derived from the best fit of the models. Three isochrones (1, 3, 10 Myr) are shown for comparison. The colors represent the number of stars present in each box.}
    \label{fig:HRD}
\end{figure}


The presence of an isochronal age spread in the HR diagram of the ONC has been previously reported and analyzed by several authors, with evidence of contamination from older populations \citep[e.g.][]{Prosser1994,Hillenbrand1997,Slesnick2004,Beccari2017, Alves2012}. In particular, 
\cite{Bouy2014} determined the presence of a large foreground population towards the Orion A molecular cloud, containing several distinct subgroups with an age spread in the range $5-10$ Myr. Still, the fraction of foreground stars contaminating our survey area centered on the Trapezium should be small with  only about $\sim 15\%$ of the WFC3-IR sources lying in the region of significant overlap between the cluster and not-cluster members, where the Bayesian's membership analysis of \cite{Robberto2020} has larger uncertainty. The same should be true for the ACS survey given, the very similar footprint.
Of the 1049 sources part of the reference catalog, only $\sim 3\%$ of them have an age estimated between 5 to 10 Myr and  $\sim 80\%$ have a distance $\lesssim 400$ pc, so it is quite possible that a number of them belong to a foreground population even if photometry shows that they are compatible with real cluster members.

\begin{deluxetable*}{|c|cccc|cc|cc|ccc|}
\tabletypesize{\footnotesize}
\setlength{\tabcolsep}{3pt}
\tablecaption{\textit{Minimum (top) and maximum (bottom) detectable mass, in solar units, as a function of filters and extinction. Filters are separed in four blocks for the ACS, WFC3/IR, NICMOS-3 and WFPC2 cameras, in the order.}
\label{Tab:mass_thersolds}}
\tablehead{
\colhead{A$_v$} & \colhead{m435} & \colhead{m555} & \colhead{m775}  & \colhead{m850} & \colhead{m130} & \colhead{m139} & \colhead{m110} & \colhead{m160} & \colhead{m336} & \colhead{m439}  & \colhead{m814}  
}
\startdata
0  & 0.025 & 0.016 & 0.004 & 0.002 & 0.003 & 0.003 & 0.009 & 0.003 & 0.152 & 0.099 & 0.027 \\
1  & 0.034 & 0.022 & 0.005 & 0.002 & 0.003 & 0.003 & 0.011 & 0.003 & 0.468 & 0.144 & 0.034 \\
2  & 0.046 & 0.030 & 0.006 & 0.003 & 0.003 & 0.003 & 0.013 & 0.003 & 0.942 & 0.328 & 0.040 \\
3  & 0.081 & 0.037 & 0.008 & 0.003 & 0.004 & 0.004 & 0.016 & 0.004 & 1.367 & 0.612 & 0.049 \\
4  & 0.118 & 0.048 & 0.010 & 0.004 & 0.005 & 0.004 & 0.020 & 0.004 & 1.367 & 1.050 & 0.068 \\
5  & 0.237 & 0.076 & 0.013 & 0.005 & 0.005 & 0.005 & 0.024 & 0.005 & 1.367 & 1.367 & 0.095 \\
6  & 0.494 & 0.108 & 0.017 & 0.007 & 0.006 & 0.006 & 0.029 & 0.005 & 1.367 & 1.367 & 0.114 \\
7  & 0.826 & 0.158 & 0.022 & 0.008 & 0.007 & 0.006 & 0.034 & 0.006 & 1.367 & 1.367 & 0.153 \\
8  & 1.262 & 0.319 & 0.028 & 0.010 & 0.008 & 0.007 & 0.041 & 0.006 & 1.367 & 1.367 & 0.250 \\
9  & 1.367 & 0.510 & 0.034 & 0.013 & 0.009 & 0.008 & 0.046 & 0.007 & 1.367 & 1.367 & 0.405 \\
\hline
\hline
0  & 0.296 & 0.104 & 0.102 & 0.102 & 0.501 & 0.456 & 0.502 & 0.201 & 1.196 & 1.087 & 0.532 \\
1  & 0.548 & 0.143 & 0.129 & 0.124 & 0.615 & 0.557 & 0.656 & 0.245 & 1.367 & 1.367 & 0.748 \\
2  & 0.929 & 0.278 & 0.187 & 0.160 & 0.765 & 0.657 & 0.873 & 0.294 & 1.367 & 1.367 & 1.057 \\
3  & 1.337 & 0.474 & 0.312 & 0.241 & 0.930 & 0.792 & 1.112 & 0.357 & 1.367 & 1.367 & 1.337 \\
4  & 1.367 & 0.676 & 0.478 & 0.359 & 1.115 & 0.932 & 1.318 & 0.421 & 1.367 & 1.367 & 1.354 \\
5  & 1.367 & 1.014 & 0.657 & 0.509 & 1.283 & 1.094 & 1.345 & 0.518 & 1.367 & 1.367 & 1.354 \\
6  & 1.367 & 1.348 & 0.946 & 0.684 & 1.343 & 1.248 & 1.345 & 0.605 & 1.367 & 1.367 & 1.354 \\
7  & 1.367 & 1.365 & 1.261 & 0.943 & 1.343 & 1.345 & 1.345 & 0.721 & 1.367 & 1.367 & 1.354 \\
8  & 1.367 & 1.365 & 1.355 & 1.222 & 1.343 & 1.345 & 1.345 & 0.863 & 1.367 & 1.367 & 1.354 \\
9  & 1.367 & 1.365 & 1.355 & 1.349 & 1.343 & 1.345 & 1.345 & 0.973 & 1.367 & 1.367 & 1.354 \\
\enddata
\end{deluxetable*}

In any case, regardless of the level of contamination by other populations, uncertainties in the measurements, presence and orientation of disk, spots, variability, anomalous extinction,  etc. can all contribute to a spread in luminosity and in turn to the observed age spread \citep{Reggiani2011}. 
Some of these effects may be associated with photometric variability.
Star spots, in particular, modulate the luminosity of a star \citep{Gully2017,Gangi2022} and
low mass young stars show variability that can be attributed to large spots created by strong magnetic fields \citep{Johns1996}. Besides variability, models show that stars in the mass range $0.1 - 1.12$ \Msun with $\gtrsim 50\%$ of their surface covered by spots have radii inflated by a factor of $10\%$ during the PMS stage \citep{Somers2015}, and between heavily spotted and spotted-free models the luminosity can change by a factor of 2. 

With all these caveats, we observe a peak in the age distribution with a median value at $\sim 1.1 \pm 0.1$ Myr. About $\sim 88\%$ of the sources are younger than 2 Myrs, while only $\sim 3\%$ of the total sources have an age estimate above 5 Myrs. Our results are consistent with a single major event of star formation in the ONC.

In Figure \ref{fig:HRD} we present the Hertzsprung–Russell Diagram (HRD) for the reference catalog of 1049 cluster sources in the form of Hess diagram. 
We bin them in boxes of $\sim 57$ K by $\sim 0.076$ log$_{10}$\Lstar 
and color-code their number according to the scheme presented in the figure. 
The  1, 3 and 10 Myr isochrones are also shown for comparison (see Section \ref{Sec:Models}).

In comparison to the previous HRDs of the ONC \citep[e.g.][]{DaRio2012}, our new HRD appears well constrained by the 1-3~Myr isochrones, with very few sources up to 10 Myr.
There is still a scatter in luminosity for effective temperatures in the range of $4000 -3000 K$ (i.e., the region dominated by our class A sources). When \Teff decreases below ~$\sim 3000 K$, our solutions rely on fewer filters and thus the priors have an increasingly stronger weight. Besides, the range of luminosities spanned by the different isochrones becomes narrower for $\Teff\lesssim 3000 K$. Both factors, the first observational and the second theoretical, contribute to the strong concentration of the sources around the $1-3$ Myr isochrones. 

\subsection{Empirical Initial mass function}
\label{sec: Empirical Initial mass function}

\begin{figure}[!t]
    \centering
    \includegraphics[width=0.45\textwidth]{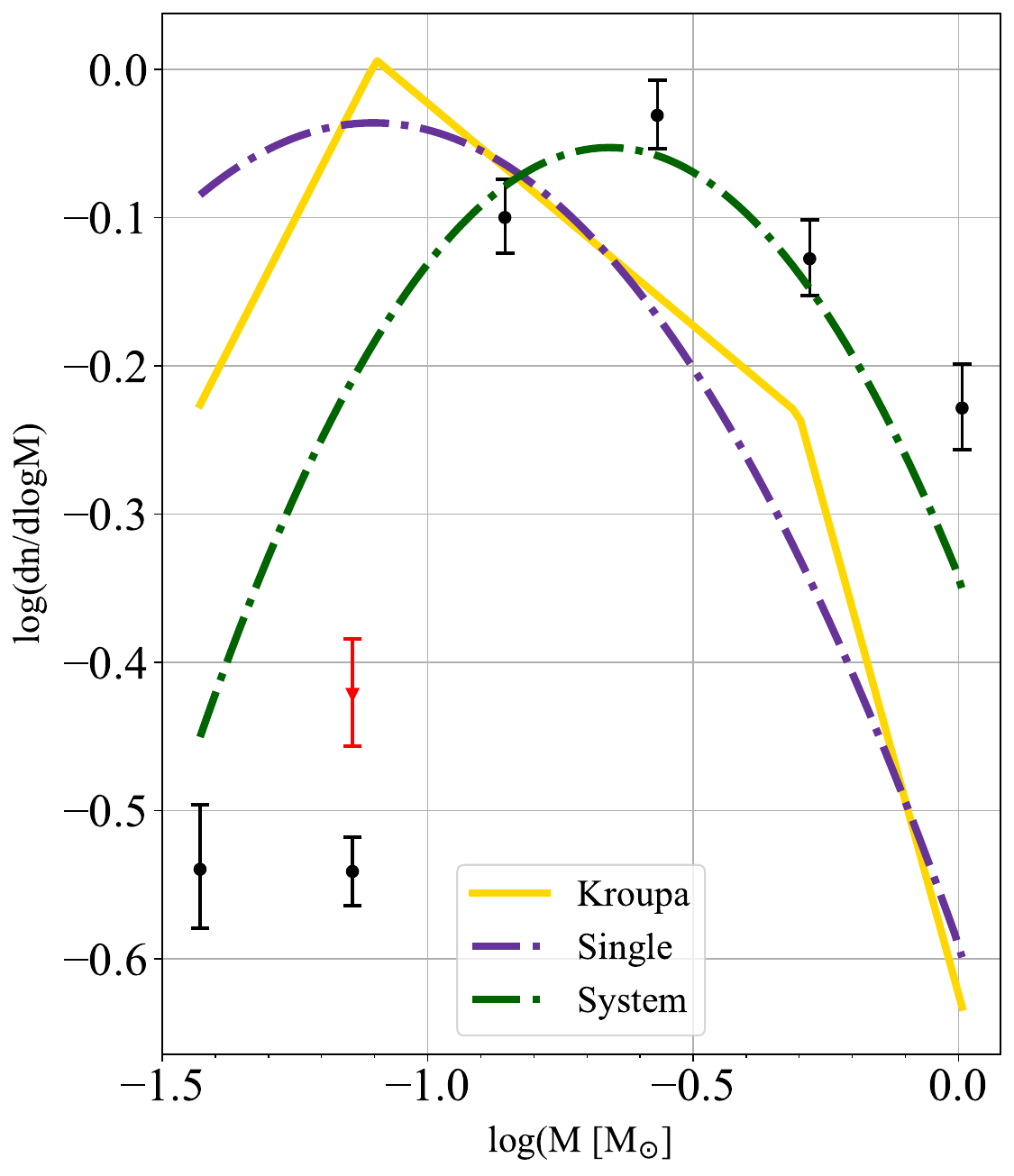}
    \caption{Comparison between the derived ONC mass distribution (black) and the canonical IMFs: Kroupa (yellow), Chabrier singles (blue) and systems (green).
    Each black dot show the average number of stars in each bin obtained form all the mass distribution histograms, with the error bars including the 1 sigma spread in each bin and the relative Poisson noise}. The red marker show the correction obtained adding \cite{DaRio2012} sources in the mass bin 0.049 - 0.095 \Msun. 
    \label{fig:IMF_sel}
\end{figure}

\begin{figure*}[!t]
    \centering
    \includegraphics[width=0.45\textwidth]{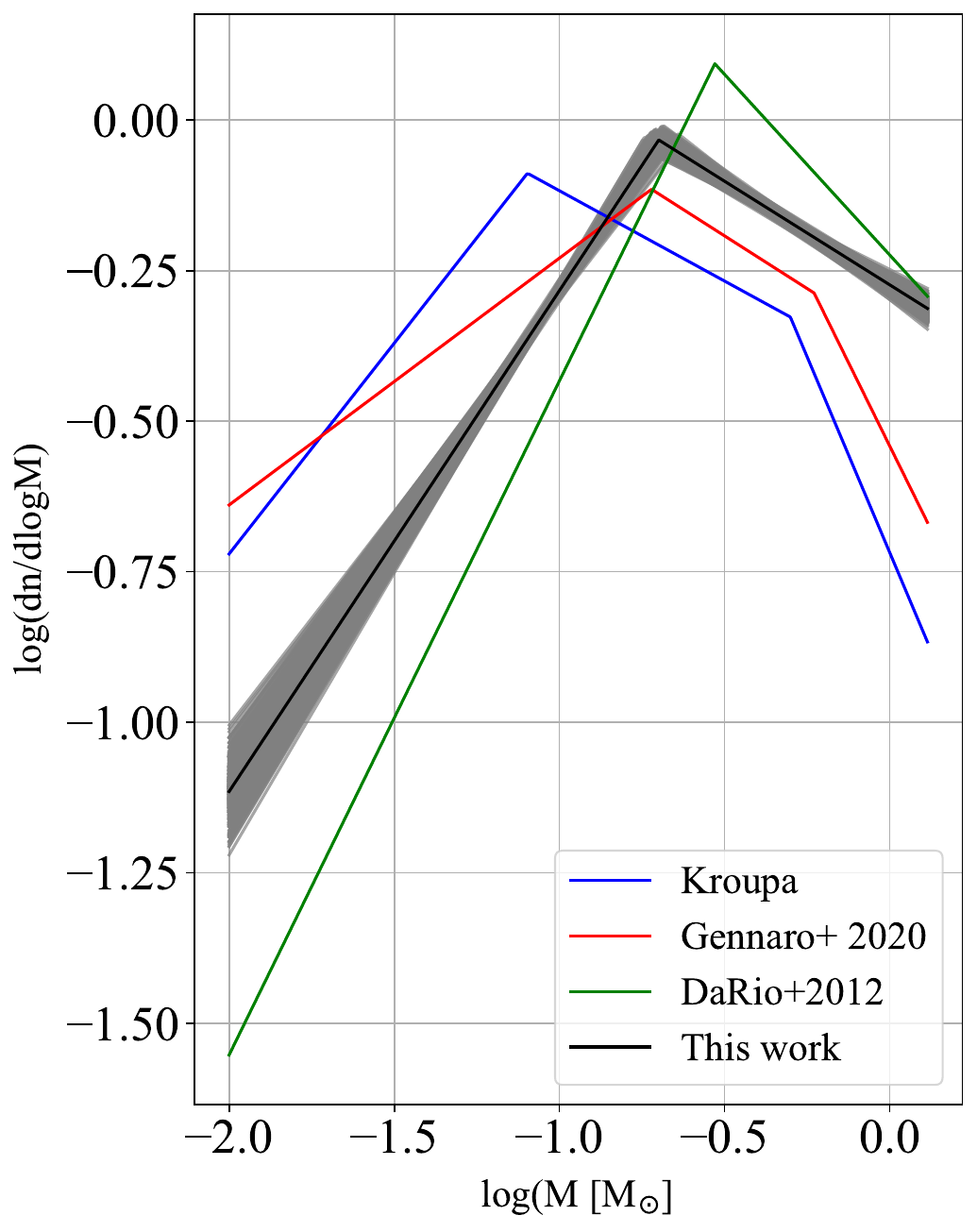}
    \includegraphics[width=0.45\textwidth]{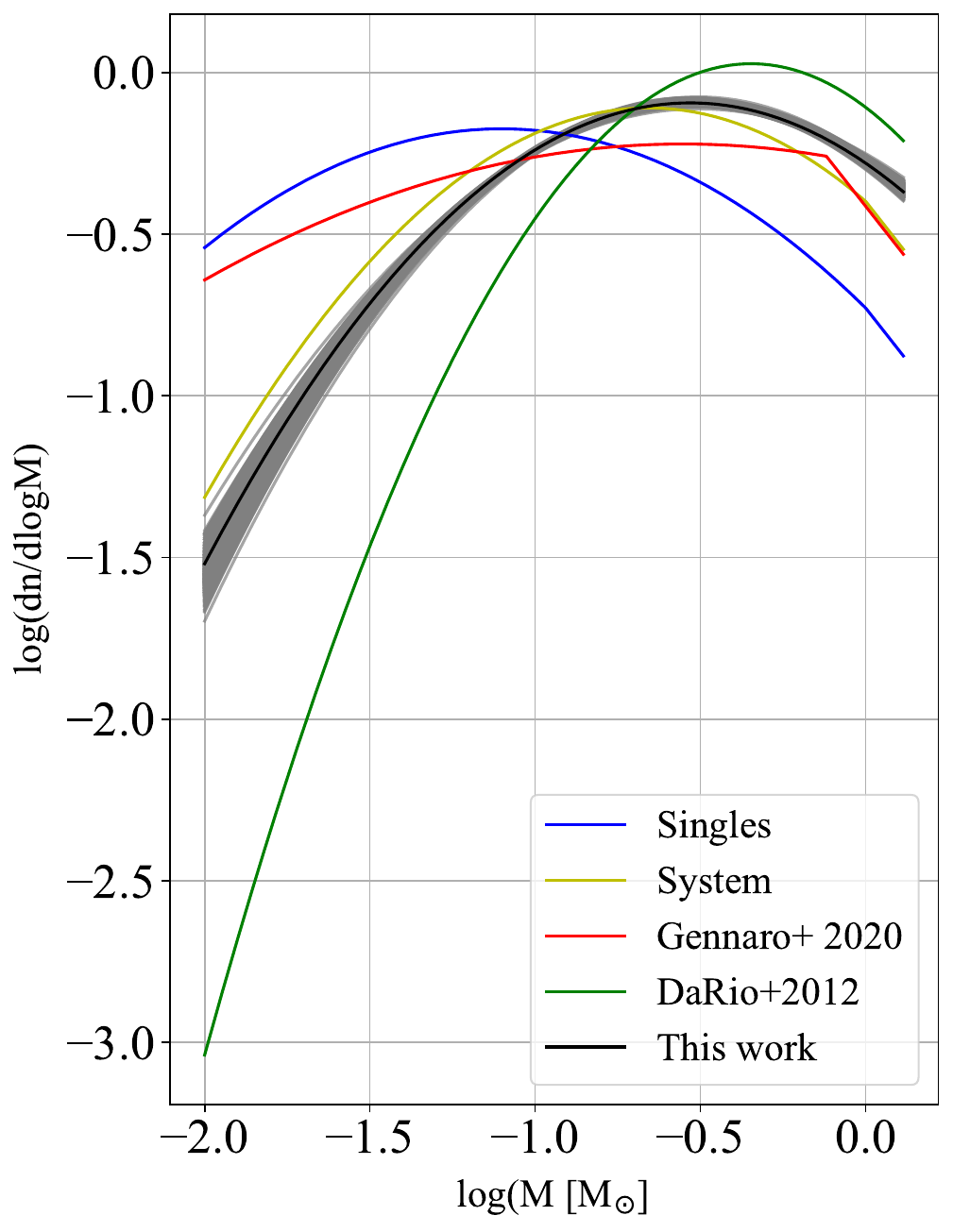}
    \caption{Comparison between the empirical ONC IMF (black) and from 500 draws from the posteriors (gray) and the canonical IMFs: Kroupa (left), Chabrier singles and systems (right). Also shown as reference in both plots the IMFs obtained by \cite{DaRio2012} and \cite{Gennaro2020}.
    \label{fig:IMF_sel2}} 
\end{figure*}

\begin{deluxetable*}{cccccc}
\tabletypesize{\footnotesize}
\setlength{\tabcolsep}{15pt}
\tablewidth{10pt}
\tablecaption{\textit{Summary of the results for the broken power law model. $\alpha_0$, $\alpha_1$ and $\alpha_2$ represent the low, intermediate and high mass slopes, while m$_{lm}$ and m$_{mh}$ represent the low-to-mid and mid-to-high transition masses.}
\label{Tab:MCMCfit1}}
\tablehead{
\colhead{Authors} & \colhead{$\alpha_{0}$} & \colhead{$\alpha_{1}$} & \colhead{$\alpha_{2}$} & \colhead{m$_{lm}$} & \colhead{m$_{mh}$}}
\startdata
        This work           &   -0.17$^{+0.04}_{-0.04}$ &   -1.34$^{+0.03}_{-0.02}$ &   -                       & 0.20$^{+0.01}_{-0.01}$ & -\\
        \cite{Kroupa2001a}  &   -0.3$^{+0.7}_{-0.7}$    &   -1.3$^{+0.5}_{-0.5}$    & -2.3$^{+0.3}_{-0.3}$      & 0.08                   &  0.5 \\
        \cite{DaRio2012} (BCAH98)   &   -0.12$^{+0.90}_{-0.90}$ &   -1.6$^{+0.33}_{-0.33}$ & -                         & 0.29                   &  -\\
        \cite{Gennaro2020}  &   -0.59$^{+0.06}_{-0.06}$ &   -1.35$^{+0.33}_{-0.35}$    & -2.11$^{+0.27}_{-0.37}$   & 0.19$^{+0.06}_{-0.05}$ &  0.59$^{+0.24}_{-0.17}$ \\
\enddata
\end{deluxetable*}

\begin{deluxetable*}{ccccc}
\tabletypesize{\footnotesize}
\setlength{\tabcolsep}{15pt}
\tablewidth{10pt}
\tablecaption{\textit{Summary of the results for the log-normal model. m$_c$ represent the characteristic mass, while $\sigma$ is the width of the log-normal, m$_{hm}$ is the transition mass and $\alpha_{hm}$ is the slope of the power-law for the high-mass regime.}
\label{Tab:MCMCfit2}}
\tablehead{
\colhead{Authors} & \colhead{m$_c$} & \colhead{$\sigma$} & \colhead{m$_{hm}$} & \colhead{$\alpha_{hm}$}}
\startdata
        This work           &   0.30$^{+0.01}_{-0.01}$ &   $0.57^{+0.01}_{-0.01}$ &   -  & - \\
        \cite{Chabrier2003}(Single)  &   0.079$^{+0.0016}_{-0.021}$    &   0.69$^{+0.01}_{-0.05}$    & 1 & -2.3 \\
        \cite{Chabrier2003}(System)  &   0.22 &   0.57    & 1 & -2.3                    \\
        \cite{DaRio2012} (BCAH98)    &   0.45$^{+0.02}_{-0.02}$ &   0.44$^{+0.05}_{-0.05}$ & - &  -\\
        \cite{Gennaro2020}  &   0.28$^{+0.12}_{-0.08}$ &   1.04$^{+0.15}_{-0.13}$    & 0.76$^{+0.24}_{-0.20}$   & -2.30$^{+0.31}_{-0.32}$ \\
\enddata
\end{deluxetable*}

Deriving an accurate Initial Mass Function (IMF) for a rich, young stellar cluster like the ONC allows in principle to discern its variations vs. e.g., radial distance from the center, isochronal age, binarity. To achieve this goal, one must obtain a complete, extinction-limited list of reliable mass values. In the case of our dataset,  the stellar masses have been  estimated combining different filter sets, and within each of them one will always find two bandpasses constraining the lowest and highest mass probed,  for any given exposure time, extinction, and assumed isochrone. 
Owing to the spread of extinction values toward the region (see Figure \ref{fig:Hists_all}) and the non uniformity of exposure times even within a single filter due e.g. to the the overlaps between adjacent frames, 
it is nearly impossible to derive homogeneous values of the limiting mass and therefore an extinction limited sample from our master catalog. 
In practice, our data analysis strategy aimed at combining a diversity of information, both priors and datasets, in order to determine the most accurate parameters of each source, is less than optimal if one wants to derive global parameters, such as the IMF, that require a homogeneous sample free from any selection bias. On the other hand, it is worth to assess our mass distribution since our methods allows retrieving sources that may be systematically rejected in more 
selective, controlled catalogs.
To illustrate the range of masses probed by the different filter combinations, we show in the top part of Table~\ref{Tab:mass_thersolds} how the minimum mass reached in each filter depends on the extinction, and similarly the lower portion shows the corresponding mass at saturation limit. Building this table we have adopted our reference isochrone and the typical exposure times of the surveys, a $\sigma \sim 0.1$ uncertainty for the faint magnitude limits, and neglected the narrow-band \Ha\, passbands of WFPC3 and ACS as they may be strongly affected by accretion or circumstellar emission. The table shows how high extinction values push low-mass sources increasingly out of reach, while high-mass stars become increasingly measurable as they fall below the saturation limits.

Therefore, a filter set composed by our ACS bandpasses will be generally limited at the low-mass limit by the bluest filter, F435W, as it is unable to reliably detect masses $M<0.025\Msun$ even when \Av= 0.
Viceversa, the upper mass limit will be determined by the wide-band F805LP filter, cutting off masses $M>1.345\Msun$ even with \Av = 9.

Considering all these caveats on the mass, we down-selected the \textit{expanded catalog} of cluster members to 1428 sources, from the initial 1546 sources (see Section \ref{sec:Membership selection}), and derived an ONC mass distribution as follows. 
Taking full advantage of our MCMC calculations, instead of adopting just the final solution for the mass of each source, we consider the last one hundred values determined by each walker as it approaches convergence.  This provides us with a rather rich and robust statistical distribution of all parameters, mass in particular, compatible with the data and the priors. 
Then, extracting randomly for each source one value from its mass distribution, we produce a particular realization of the mass distribution in ONC. We repeat this process a hundred times, each iteration providing us with a slightly different IMF, that once averaged over the same constant bins produce the mass distribution represented by the black markers in Figure \ref{fig:IMF_sel}. The error bars indicate the $1\sigma$ spread of the data in each bin, the only anomaly being the dip observed in the bin at about 0.07 \Msun. This is most probably a selection effect. Stars falling in this bin are typically at the sensitivity limit of the the bluer ACS, and WFPC2 filters, yet massive enough to be poorly classified using their shallow WFC3 1.4~$\mu$m absorption band (see Section \ref{sec:Observation and Data Reduction} for more details). The two effects combined may cause MCMC to provide degenerate or erroneous solutions. The aforementioned clustering of sources foreground objects toward the center of the cluster confirms that we do not account for all cluster members in the central region. On the other hand, our estimates --both mass and distance, i.e. membership-- are more robust for larger masses, where we generally have a more complete SED, and for smaller masses where the WFC3 1.4~$\mu$m absorption band becomes prominent.
The deficit of stars can be reconciled by including the stars that \cite{DaRio2012} classified as belonging to this mass bin. This operation adds back 39 sources in the mass bin between 0.049 and 0.095 \Msun, obtaining the point represented by the red triangle in Figure~\ref{fig:IMF_sel} and a final catalog of 1467 total sources.

It is well established that for the low-mass stars and BDs (which is the region of the mass spectrum most relevant for our study), the IMF can be generally approximated with a shallower power law \citep[$\xi(log \: m) \propto m^{-\alpha}$; ][]{Kroupa2001a} or a log-normal function \citep[$\xi(log \: m) \propto e^{-(log \: m - log \: m_c)^2/2\sigma^2}$; ][]{Chabrier2003}. Adopting a Bayesian approach and an MCMC algorithm, we use both forms to fit our measured mass distribution in the mass range 0.025 - 1.34 \Msun. The resulting distributions are shown in figure \ref{fig:IMF_sel2}, with the best fit parameters presented in Tables \ref{Tab:MCMCfit1} and \ref{Tab:MCMCfit2}. For reference, are also reported the values from the canonical IMFs (Kroupa - \citealp{Kroupa2001a}, Chabrier singles and systems - 
 \citealp{Chabrier2003}), as well as the ONC IMFs obtained by \cite{DaRio2012} and \cite{Gennaro2020}.

When we compare the results for the broken power-law case from Table \ref{Tab:MCMCfit1}, our low-mass slope, $\alpha_0 = -0.17$ is compatible within the 1 $\sigma$ with \cite{DaRio2012} measurement. The discrepancy vs. \cite{Gennaro2020} is due to our data not accounting for unresolved multiple systems with separation closer than $0.1"$ \citep[or $\sim 40 AU$ at the distance of the ONC;][]{Strampelli2023}, while \cite{Gennaro2020} leave the binary fraction free to vary in their bayesian model. 
The transition mass between the low and mid ranges, $m_{lm} = 0.20$ \Msun, falls between the value predicted by \cite{Kroupa2001a} and the one observed by \cite{DaRio2012}, and is also in very good agreement with the prediction from \cite{Gennaro2020}. For the intermediate-mass slope, $\alpha_1$ = -1.34, we are in very good agreement with both predictions from \cite{Kroupa2001a} and \cite{Gennaro2020}. On the other hand, we observe a shallower $\alpha_1$  slope compared to the one observed by \cite{DaRio2012}, but their large uncertainty makes the results still compatible.

For the log-normal case (Table \ref{Tab:MCMCfit2}), we determine a peak mass compatible with \cite{Gennaro2020} IMF, but with smaller $\sigma$ (our distribution is narrower). Similar to the power-law case, our derived mass function is indeed expected to be more similar to a “system” mass function as we do not try to resolve close-in companions in our data. Compared to \cite{DaRio2012}, we observe a smaller peak mass and a larger distribution (higher $\sigma$ value).

Over all, for masses $\gtrsim 0.1$ \Msun our empirical IMF is in good agreement with a \cite{Chabrier2003} System IMF.

\subsection{Accretion rates in the ONC}\label{sec:Accretion rates in the ONC}

\begin{figure*}[!t]
    \centering
    \includegraphics[width=0.45\textwidth]{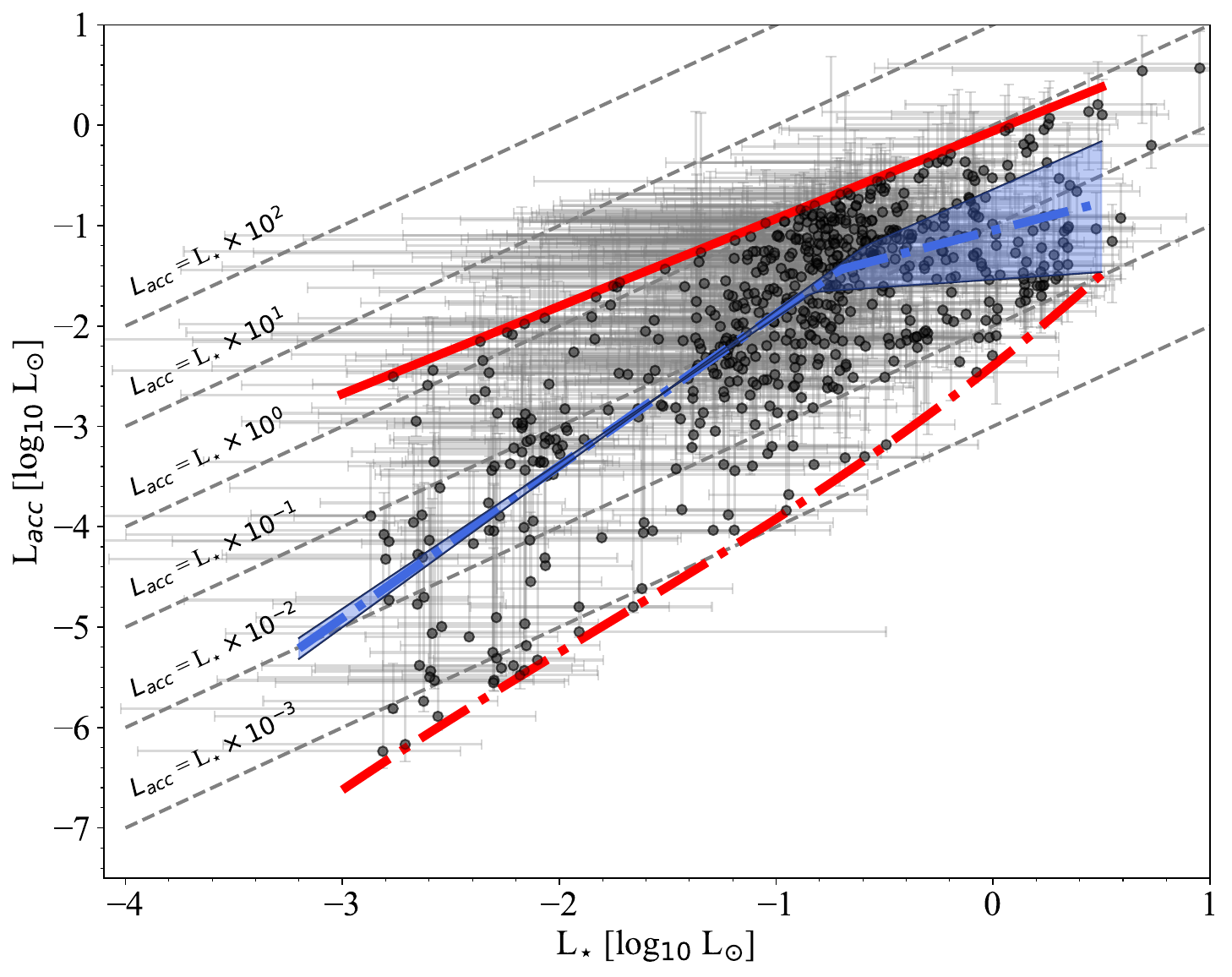}
    \includegraphics[width=0.45\textwidth]{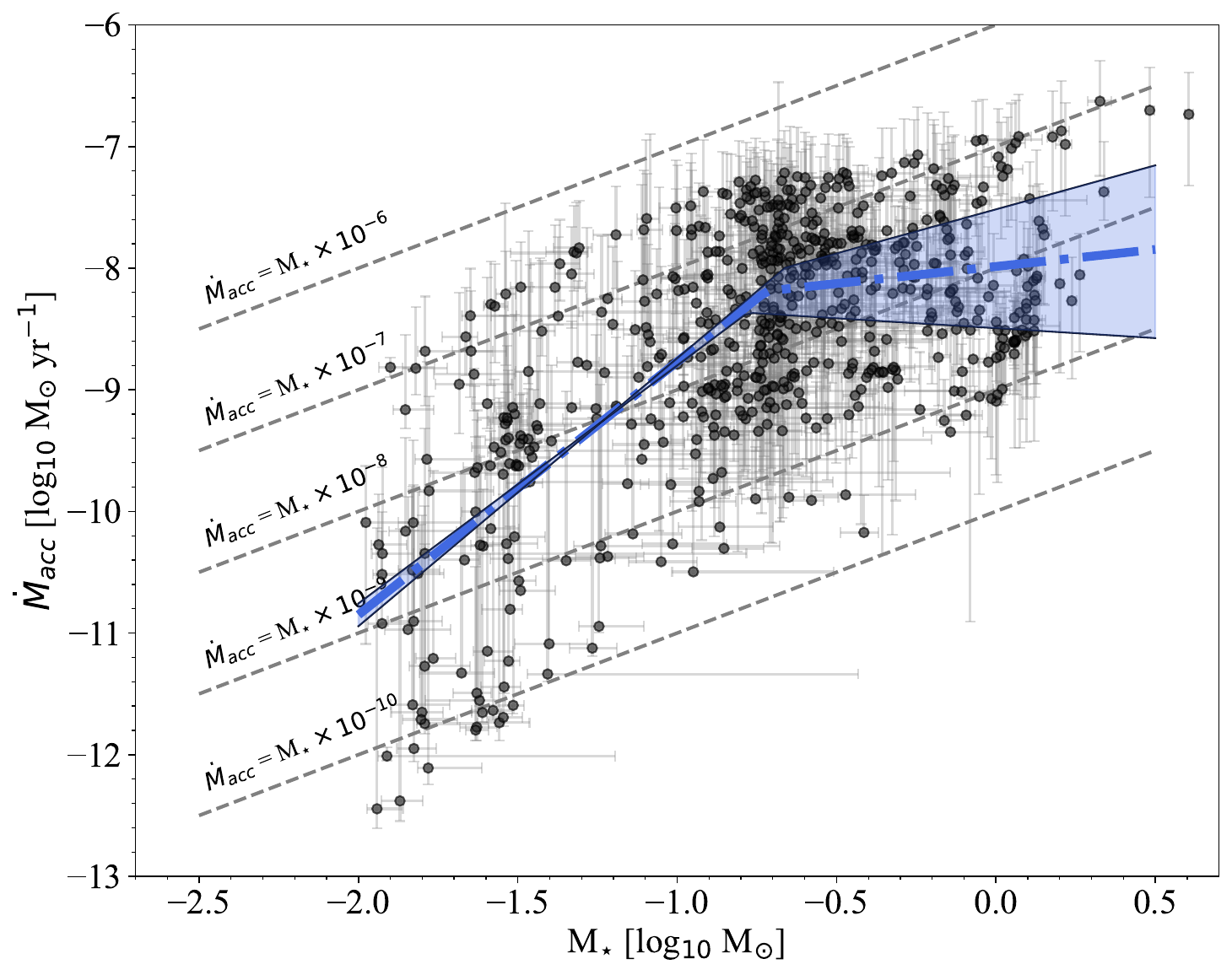}    
    \caption{
    Accretion luminosity (\Lacc) as a function of stellar luminosity (\Lstar) for ONC sources with relative error bars. Also shown are the limits derived due to chromospheric emission for a 1 Myr old object (lower dash-dotted red curve) and the limit above which we expect the majority of sources to be proplyd (continuous red line), as explained in the text. Right: Similar to left but for the Mass accretion rate (\dMacc) as a function of stellar mass (\Mstar) for ONC sources. The thick blue dashed segmented line represents the best fit obtained from the data using a piecewise linear function (equation \ref{eq:logLacc_fitt}). The colored area provides a relative visualization of the errors on the fit. }
    \label{fig:accretion_plots}
\end{figure*}

In Figure \ref{fig:accretion_plots}, left panel, we present the relation between \Lacc and \Lstar for our selected cluster sources. Similar to Sec. \ref{sec:Hertzsprung–Russell Diagram }, we apply a stronger selection to reference catalog including only the sources with AR $> 50\%$. The upper solid red line limits the area above which the majority of sources coincide with proplyds, according to the catalog by \cite{2008AJ....136.2136R}, making the estimate of \Lacc unreliable. 
The bottom dash-dotted red curve instead marks the stellar photospheric limit \citep[see Eq. 1 in ][]{Manara2017b} for 1 Myr sources, below which the chromosphere is as bright or brighter than the accretion excess and therefore prevents disentangling the \Lacc contribution. 

Selecting only the the reference catalog between these two lines, and applying the stronger cluster selection discussed above, we can build a plot showing the \dMacc - \Mstar relation for 631 sources (Figure \ref{fig:accretion_plots}, right panel). 
Similar to the cases of Chamaleon I and Lupus \citep{Alcala2017,Manara2017a},  \dMacc appear to increase with \Mstar. The scatter is large, but the relation appears steeper for low-mass objects, suggesting that one may consider either a single or a broken power-law relation. We tested which approach provides a better description of our data (see Appendix \ref{app:Power-law Vs Broken Power-law fit}) finding that a segmented linear relation (corresponding to a broken power law in the linear scale) is preferable over a linear fit. With the already mentioned caveats about saturation and biases, we can provide for $0.01$ \Msun $\lesssim$ \Mstar $\lesssim 0.7$ \Msun the following relations:

\begin{equation}
    \begin{split}
        \logLacc= - (0.40\pm0.09) + (1.46 \pm0.06)\: \logLstar  \\  \textrm{if} \ \logLstar \leq -0.63 \\
        \logLacc= - (0.99\pm0.09) + (0.42\pm0.40)\: \logLstar  \\  \textrm{if} \ \logLstar > -0.63
    \end{split}
    \label{eq:logLacc_fitt}
\end{equation}
while for \dMacc we obtain:
\begin{equation}
    \begin{split}
        \logdMacc=- (6.78\pm0.12) + (1.98\pm0.10) \: \logMstar  \\  \textrm{if} \ \logMstar \leq -0.68 \\
        \logdMacc=- (7.92\pm0.24) + (0.13\pm0.41) \: \logMstar  \\  \textrm{if} \ \logMstar > -0.68
    \end{split}
    \label{dM_acc_fitt}
\end{equation}

These relations are shown in Figure \ref{fig:accretion_plots} as a blue dashed line. The filled area around each fit provides a visualization of the uncertainties associated with each fit. 

A broken power-law relation for both \Lacc and \dMacc distributions is a new result compared to previous studies of accretion in young star-forming region \citep[e.g.; ][]{Manara2017b}, and in particular for this cluster \citep{Manara2012}, where a linear relation was generally preferred as there was not enough evidence to distinguish between them.
For both parameters, the distributions appear strongly dependent on  \Mstar (or \Lstar) below a \Mstar $\lesssim 0.21$ \Msun (or \Lstar $\lesssim 0.23$ \Lsun). After, the trend becomes flatter, indicating that the \dMacc (\Lacc) remains more or less constant as \Mstar (\Lstar) increase. However, there is a large spread of values (about 2 orders of magnitudes) for both distributions.

Our results on the \Lacc - \Lstar can be compared with the relations obtained in other SFRs like e.g. $\rho-$Ophiuchi \citep[][]{Natta2006}, $\sigma-$Orionis \citep[][]{Rigliaco2011}, Chamaeleon I \citep[][]{Manara2016a,Manara2017a} and Lupus \citep[][]{Alcala2014,Alcala2017}. If we only consider the region where \Lstar $\lesssim 0.23$ \Lsun (before the braking point) we find a good agreement between our portion of the fit and the one obtained for Chamaeleon I, while we find a more loose match (within $3\sigma$) with the other SFRs. Similarly, the slope of the \dMacc $-$ \Mstar relationship is in broad agreement with other SFRs \citep[e.g. Chamaleon I, Lupus and NGC 2264: ][]{Alcala2017,Manara2017a,Venuti2014}, where the data seem  to better support a double power-law fit with lower mass stars having a more rapid decrease of \dMacc compared to higher mass stars (even though the authors could not completely rule out the linear dependence). 
In general, when comparing results between different star forming region, age differences should be taken into account. Both the small variation observed with the ages and the high uncertainties related to age estimations itself \cite[e.g.,][]{Soderblom2014} will make any detailed analysis very difficult to perform.

The slope of the \dMacc relation can also be compared with theoretical models and be used as a proxy to evaluate the main processes behind the dispersal of protoplanetary disks \citep{Clarke2006}. Typical values suggest power laws with exponent $\sim 1.6 - 2$ and spread of $\sim 1-2$ dex.
\citep{Alcala2014,Alcala2017,Manara2016b,Manara2017b,Hartmann2016}. 
Recent works, however, show that accretion variability alone is not enough to explain the spread observed in \dMacc measurements \citep{Fischer2022,Flaischlen2022}  that must therefore be also related to  physical processes such as disk evolution. Is still not yet clear in what measure the \dMacc $\propto$ $M_{*}^2$ relation is due to the disk evolutionary process or if it reflects how the initial conditions scale with stellar mass \citep[][and references therein]{Manara2022}. 
For example, \cite{Dullemond2006} were able to produce a steep \dMacc - \Mstar relation resulting from the imprint of the initial conditions for the formation of the star-disk system. A simple model of disk formation and evolution from collapsing cores provides a \dMacc $\propto$ \Mstar$^{1.8}$ relation, provided that cores of all mass have a similar distribution of rotation periods. 

The ONC environment is  characterized by strong UV flux that affects the structure and evolution of the disks closer to the central OB stars \cite{2022EPJP..137.1132W}. 
However, from our finding, at least for low-mass stars, the \dMacc - \Mstar relation is remarkably similar to the one observed in low-mass SFRs suggests that external photoevaporation does not play a significant role on the majority of young circumstellar disks. The slope of the relation, therefore, may be largely determined by the initial conditions set at the onset of the star formation process, which may be quite similar between regions that eventually form clusters of different sizes \citep{Manara2022}. 

Photoevaporation by the disk central star can also play a role in the \dMacc - \Mstar slope. X-ray-driven photoevaporation produces a slope between 1.6 - 1.9 \citep{Ercolano2014} while UV-driven photoevaporation provides a value of $\sim 1.35$ \citep{Clarke2006}. 
In this context, the result we obtain for \Mstar $\lesssim 0.21$ \Msun is probably more in line with the X-ray-driven scenario rather than the UV one, even though both the X-ray-driven and collapsing cores scenario provide shallower slopes compared to ours.

A consequence of a bimodal distribution on the \dMacc-\Mstar plane is a different evolutionary timescale for disk accretion around stars with different masses: i.e., disks around stars with \Mstar $\lesssim 0.2$ \Msun will evolve faster toward lower values of \dMacc. The surveys in $\sigma-$Orionis \citep{Rigliaco2011}, in the same ONC \citep{Manara2012}, as well as the spectroscopic survey of L1641 \citep{Fang2013}, arrived to similar conclusions, even though we need to point out that a faster evolution for lower-mass stars is opposite to predictions by \cite{Alexander2006}, who postulated that the viscous timescale increases with stellar mass to explain the \dMacc $\propto$ \Mstar$^{2}$ relation. 

A reason behind the bimodal distribution can be found in two different physical regimes at play in the early stages of disk formation and evolution \citep{Vorobyov2009}. These authors suggested that disk self-gravity will drive large accretion torques for stars more massive than $\sim 0.2$ \Msun soon after formation. This in turn will leave less material available in the Class II phase and therefore a lower accretion rate which will flatten the relationship. For lower mass stars instead, self-gravity is less important and disks evolve more viscously. In particular, the least-squares fit their model data provide as an exponent:
\begin{equation}
    \begin{split}
        n = 2.9 \pm 0.5 \ \textrm{if} \ \Mstar < 0.2 \Msun \\
        n = 1.5 \pm 0.1 \ \textrm{if} \ 0.2 \leq \Mstar < 3 \Msun \\
    \end{split}
\end{equation}
while the corresponding fits to the observation from \citep[][and references therein]{Muzerolle2005}  for TTSs and BDs of age $0.5-3.0$ Myr  produce:
\begin{equation}
    \begin{split}
        n = 2.3 \pm 0.6 \ \textrm{if} \ \Mstar < 0.2 \Msun \\
        n = 1.3 \pm 0.3 \ \textrm{if} \ 0.2 \leq \Mstar < 3 \Msun \\
    \end{split}
\end{equation}
Remarkably, 
the \citep{Vorobyov2009} model predicts a 
a similar relation for \Mstar $\lesssim 0.21$ \Msun compared to ours with our estimate positioning itself between the model and the observational data relations. 
For  \Mstar $> 0.21$ \Msun instead, our fit is much shallower, even though we have to argue here that our selected sample is limited to \Mstar higher masses due to saturation of some or more filters (only $\sim 15\%$ of sources in our selected catalog have higher masses). On the other hand, their data reach up to 3 $\Msun$, so we are not covering quite the same mass range. 



\section{Conclusion}
\label{sec:Conclusion}
We have revisited the   \cite{Robberto2013} \textit{HST}/ACS/WFC ONC catalog in filters F435W, F555W, F658N, F775W, and F850LP to provide updated estimates of the  average magnitudes in each filter. We then combined the photometry from ACS and WFPC2 with the more recent \textit{HST}/WFC3-IR measures from \cite{Robberto2020} in filter F130N and F139M. The resulting dataset contains the photometric data derived from the two \textit{HST} Treasury programs dedicated to the ONC, spanning almost 15 years. 
We adopted a Bayesian SED fitting to derive reliable estimates of the principal stellar parameters (i.e., temperature, extinction, age, accretion, and distance) for the majority of the sources in the catalog and estimate general properties for the cluster. In particular, the three dimensional study of mass distribution for bona-fide cluster members shows that
mass segregation in the ONC extends to sub-solar masses, possibly down to $\sim 0.2$~\Msun.
From the age distribution we unveil a star formation history heavily peaked at $\sim 1.1 \pm 0.1$ Myr, with a small tail extending up to 10 Myrs. Our estimates, model dependent, do not support the idea of multiple episodes of significant star formation in the ONC. Training our dataset on the group of cluster members with known parallax, we evaluate an average distance for the cluster of $397 \pm 17$~pc, and similarly for the extinction distribution we estimate an median extinction of $4.35\pm4.10$ mag.

The inclusion of accretion in our fitting procedure allows us to test the relationship between \Lacc and \dMacc vs. the stellar parameters. If we consider only the region where \Lstar $\leq 0.16$ \Lsun, we also find good agreement between the  \Lacc - \Lstar relation for this cluster and a less massive system like e.g. Chamaeleon I, while we find a more loose agreement (within $3\sigma$) with $\sigma-$Orionis, $\rho-$Ophiuchi, and Lupus SFRs.
Overall, we find evidence for a bimodal relation for  \Lacc and \dMacc for this cluster. Both can be parameterized as broken power-laws with slopes \logLacc $1.46 \pm 0.06$ for \logLstar $\leq -0.63$ and $0.42 \pm 0.40$ otherwise, and \logdMacc $1.98 \pm 0.10$ for \logMstar $\leq -0.68$ and $0.13 \pm 0.41$ otherwise. 

The result we obtain for \Mstar $\lesssim 0.2$ \Msun support a scenario where the excess emission is dominated by X-ray-emission from the central star rather than the UV one, while external photoevaporation doesn't look like playing a significant role on the majority of young circumstellar disks. The slope of the relation, therefore, may be largely determined by the properties of the central stars, and therefore by the initial conditions set at the onset of the star formation process, rather than by the dynamical evolution of the cluster. These conditions may be quite similar between regions that eventually form clusters of different sizes \citep{Manara2022}. 


The shape of the IMF turns out to depend on the particular dataset, as the type and number of filters used to derive the source parameters is not homogeneous. This introduces selection biases that need to be carefully assessed when deriving global systems parameters such as the IMF. 
Overall, for masses $\gtrsim 0.1$ \Msun, we find our results to be compatible with a canonical Chabrier System IMF.

This work underlines the potential and limitations of multi-color photometric surveys to characterize the main physical properties of star forming regions.

\acknowledgments
The authors thank the anonymous referees for their useful comments on the manuscript.
G. M. Strampelli wants to thank the Instituto de Astrofísica de Canarias for its hospitality.  The authors thank Gaspard Duchene for the invaluable suggestions and comments. Support for Program number GO-10246 was provided by NASA through a grant from the Space Telescope Science Institute, which is operated by the Association of Universities for Research in Astronomy, Incorporated, under NASA contract NASS-26555.  JA was supported in part by a grant from the National Physical Science Consortium. GMS and AA acknowledge support from the Spanish Agencia Estatal de Investigación del Ministerio de Ciencia e Innovacion (AEI-MICINN) under grant "Proyectos de I+D+i" with references AYA2017-89841-P and PID2020-115981GB-I00.
CFM is funded by the European Union under the European Union’s Horizon Europe Research \& Innovation Programme 101039452 (WANDA). Views and opinions expressed are however those of the author(s) only and do not necessarily reflect those of the European Union or the European Research Council. Neither the European Union nor the granting authority can be held responsible for them.

\dataset\\The data presented in this paper were obtained from the Mikulski Archive for Space Telescopes (MAST) at the Space Telescope Science Institute. The specific observations analyzed can be accessed via \dataset[DOI]{https://doi.org/10.17909/rr5s-js47}

\facilities{\textit{HST} (ACS, WFC3)}
\software{Numpy \citep{Numpy}, Astropy \citep{astropy:2013,astropy:2018}, Scipy \citep{SciPy-NMeth:2020}, Matplotlib \citep{Hunter:2007}, PyKLIP \citep{Wang2015}, Pandas \citep{mckinney2010data}}

\newpage

\FloatBarrier
\clearpage
\appendix

\section{Power-law Vs. Broken Power-law fit}
\label{app:Power-law Vs Broken Power-law fit}
\setcounter{table}{0}
\setcounter{figure}{0}

\renewcommand{\thefigure}{C\arabic{figure}}
\renewcommand{\thetable}{C\arabic{table}}

\begin{figure*}[!h]
    \centering
    \includegraphics[width=0.45\textwidth]{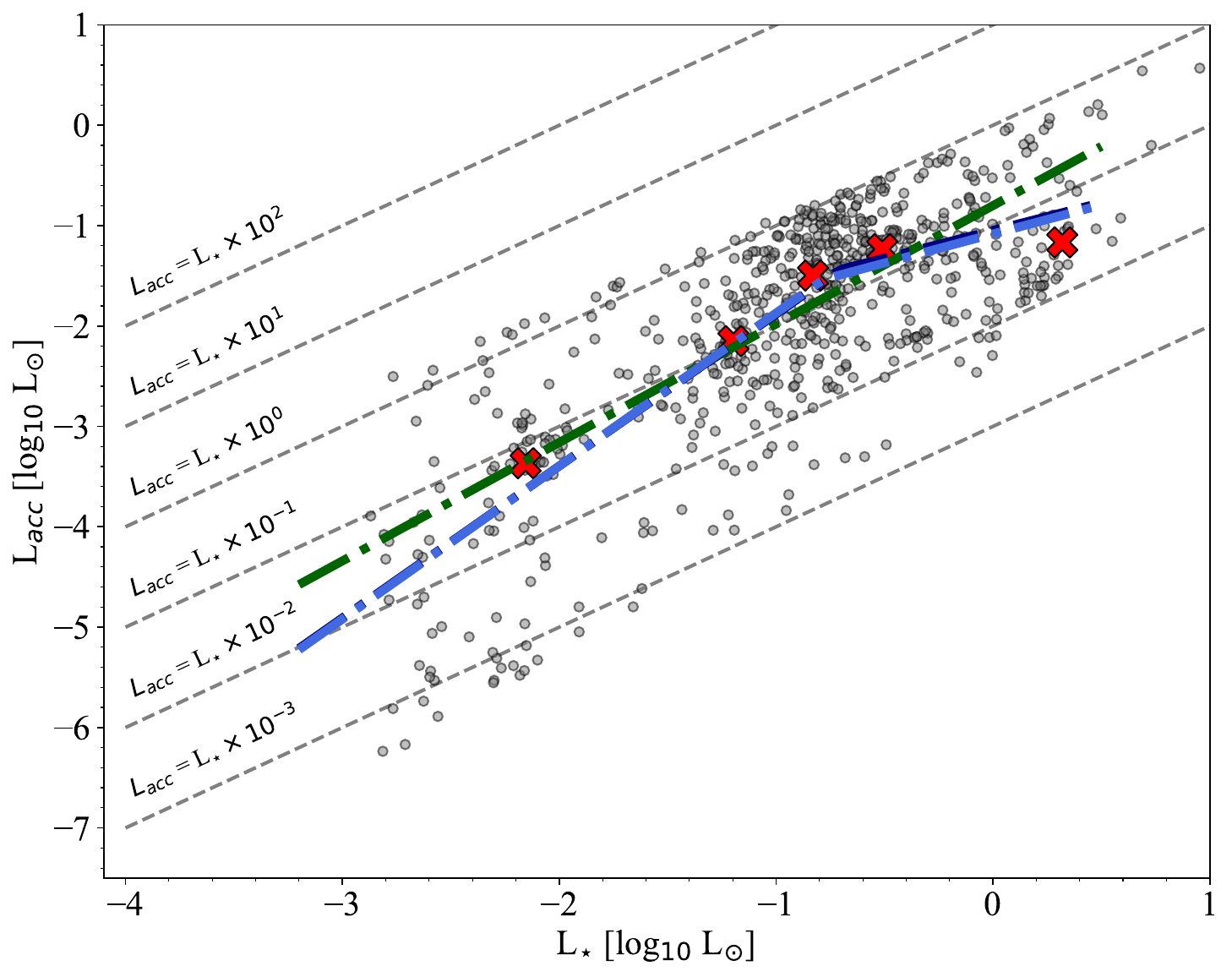}
    \includegraphics[width=0.45\textwidth]{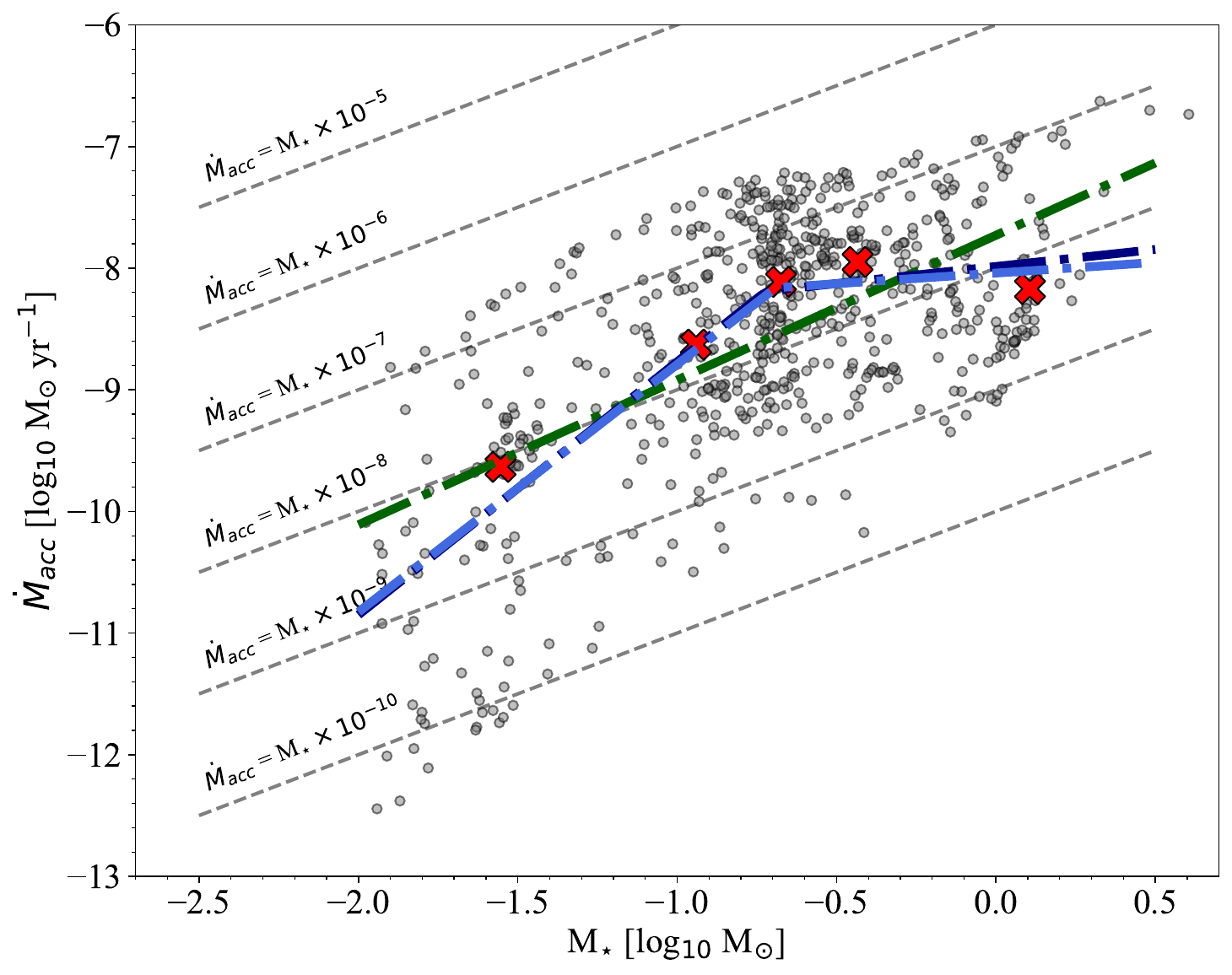}
    \caption{
    Left: Accretion luminosity (\Lacc) as a function of stellar luminosity (\Lstar) for ONC sources
    . Also shown is the fit for the linear relation (green) and bimodal relation (dark blue: selected, light blue: not selected). Right: Similar to left but for the Mass accretion rate (\dMacc) as a function of stellar mass (\Mstar) for ONC sources.
    }
    \label{fig:accretion_plots_medians}
\end{figure*}

Here we report on additional tests we performed to study the dependence of the accretion on the stellar parameters as presented in Section \ref{sec:Accretion luminosity and mass accretion rate}.
To test whether a single power-law provides a better model for our data compared to a broken power-law (corresponding to a line or segmented line in logarithmic space), we adopted an approach similar to the one presented in \cite{Manara2017a}. We started dividing both \logLacc and \logdMacc distributions in bins of an equal number of sources and we evaluate the median value for each bin (represented by the red crosses in Figure \ref{fig:accretion_plots_medians}). Both relations show a bending in their trend when approaching higher values of \Lstar or \Mstar.

Then we fitted the data with both a linear relation:
\begin{equation}
    y=\theta_0+ \theta_1 \cdot x
\end{equation}
or a segmented  line:
\begin{equation}
    \begin{split}
        y=\theta_0 + \theta_1 \cdot x \qquad  \textrm{if} \ x \leq x_c \\
        y=\theta_0 + \theta_1 \cdot x + \theta_2 \cdot [x-x_c] \qquad  \textrm{if} \ x> x_c
    \end{split}
\end{equation}
where $\theta_i$ (i=0,1,2) and $x_c$ are free parameters for the fit and y is either log(\Lacc) or log(\dMacc) and x is either log(\Lstar) or log(\Mstar), depending on the case.

We perform the linear fit using \texttt{RANSAC} \citep{Fischler1981} \texttt{scikit-learn} python module \citep{Pedregosa2011}. The fit has been performed one hundred times, where during each iteration one hundred trials were performed by the \texttt{RANSAC} routine to establish the best-fit parameters, and a final median and standard deviation have been extracted from the overall output distributions of parameters, providing the following best fit:
\begin{equation}
    \logLacc= - (0.79\pm0.13) + (1.17\pm0.10) \: \logLstar 
    \label{eq:logLacc_fitt}
\end{equation}
The same linear fit applied to the \dMacc - \Mstar relation provide instead the following best fit:
\begin{equation}
    \logdMacc=- (-7.68\pm0.14) + (1.21\pm0.14) \: \logMstar 
    \label{dM_acc_fitt}
\end{equation}
both relations are shown as green dashed line in Figure \ref{fig:accretion_plots_medians}.

The segmented line instead has been fitted using \texttt{piecewise regression} \citep{Pilgrim2021} that fit simultaneously the breakpoint positions and the linear models using an iterative method. Similar to the \texttt{RANSAC} case, a thousand trials were performed by the routine to establish the best-fit parameters, and a final median and standard deviation have been extracted from the overall distributions of parameters.
For this case, the best-fit parameters are for \Lacc are $\theta_0 = - 0.39\pm0.08$, $\theta_1 = 1.47 \pm 0.06$, $\theta_2 = -0.95 \pm 0.15$, and $x_c = -0.70 \pm 0.1$ that can be translated to: 
\begin{equation}
    \begin{split}
        \logLacc= - (0.39\pm0.08) + (1.47 \pm0.06)\: \logLstar  \\  \textrm{if} \ \logLstar \leq -0.70 \\
        \logLacc= - (1.11\pm0.08) + (0.51\pm0.14)\: \logLstar  \\  \textrm{if} \ \logLstar > -0.70
    \end{split}
    \label{eq:logLacc_fitt}
\end{equation}
while for \dMacc we obtain $\theta_0 = - 6.77\pm0.14$, $\theta_1 = 1.98 \pm 0.11$, $\theta_2 = -1.84 \pm 0.19$, and $x_c = -0.68 \pm 0.05$, or:
\begin{equation}
    \begin{split}
        \logdMacc=- (6.77\pm0.14) + (1.98\pm0.11) \: \logMstar  \\  \textrm{if} \ \logMstar \leq -0.68 \\
        \logdMacc=- (8.11\pm0.14) + (0.14\pm0.15) \: \logMstar  \\  \textrm{if} \ \logMstar > -0.68
    \end{split}
    \label{dM_acc_fitt}
\end{equation}
both relations are shown as blue dashed line in Figure \ref{fig:accretion_plots_medians}.

To compare the results of the linear and segmented models, we selected three information criteria: the $R^2$, the Akaike information criteria (AIC), and the Bayesian information criteria (BIC). 
All of them provide a statistical measure of how well the model approximates the real data, but the BIC and AIC also take into account the number of free parameters included in the model, penalizing the results accordingly. Overall, we can identify the better model by looking for the one that maximizes $R^2$ and at the same time minimize both the AIC and BIC. We assume that a difference in AIC between two models of 14 or more is enough to exclude the model with higher AIC \citep{Murtaugh2014}, while a difference in BIC of 10 or more is sufficient to exclude the model with higher BIC \citep{Riviere2016}.
For the power-law only, the \texttt{piecewise regression} algorithm also performs the Davies test (p) for the existence of at least 1 breakpoint. In general, if $p<0.05$ means reject the null hypothesis of no breakpoints  at $5\%$ significance \citep{Davies1987,Davies2022}.
\begin{table}[!t]
    \centering
    \caption{Summary of the results for the selected information criteria adopted in this work to compare the single and broken power-law models obtain from our data.}
    \begin{tabular}{c|c|c|c|c|c}
     \hline
     \hline
                    &  $R^2$ & AIC & BIC & p                        & Case\\
     \hline
     \logLacc       &   0.59 & -353 & -335 & -                      &single power-law\\
     \logLacc       &   0.62 & -409 & -391 & $4.6\times10^{-7}$     &broken power-law\\
     \hline
     \logdMacc      &   0.39 & -258 & -240 & -                      &single power-law\\
     \logdMacc      &   0.47 & -331 & -349 & $5.6\times10^{-13}$     &broken power-law\\
     \hline
    \end{tabular}
    \label{tab:information_criteria}
\end{table}

The results obtained for both models are presented in Table \ref{tab:information_criteria}, showing that the broken power-law maximizes the $R^2$ while minimizing both the AIC and the BIC for both the \logLacc and \logdMacc distributions (with a difference greater than the threshold provided above) and it can be assumed as a better descriptor of these distributions. Moreover, the broken power law always has $p<<0.05$ so the presence of a braking point is very likely. Because the second portion of the broken power-law fit (in particular for the \dMacc - \Mstar relation) seems to be driven by a cluster of massive stars with lower \dMacc closer to the right end of our distribution (i.e. sources with \logMstar $\gtrsim -0.15$ or $\sim 0.7$ \Msun), we decide to test whether the predilection for a broken power-law relation still stand if we eliminate these sources.

In this case, we obtain $\theta_0 = - 0.4\pm0.09$, $\theta_1 = 1.46 \pm 0.06$, $\theta_2 = -1.04 \pm 0.4$, and $x_c = -0.63 \pm 0.11$ that can be translated to: 
\begin{equation}
    \begin{split}
        \logLacc= - (0.40\pm0.09) + (1.46 \pm0.06)\: \logLstar  \\  \textrm{if} \ \logLstar \leq -0.63 \\
        \logLacc= - (0.99\pm0.09) + (0.42\pm0.40)\: \logLstar  \\  \textrm{if} \ \logLstar > -0.63
    \end{split}
    \label{eq:logLacc_fitt}
\end{equation}
while for \dMacc we obtain $\theta_0 = - 6.78\pm0.12$, $\theta_1 = 1.98 \pm 0.10$, $\theta_2 = -1.84 \pm 0.10$, and $x_c = -0.68 \pm 0.07$, or:
\begin{equation}
    \begin{split}
        \logdMacc=- (6.78\pm0.12) + (1.98\pm0.10) \: \logMstar  \\  \textrm{if} \ \logMstar \leq -0.68 \\
        \logdMacc=- (7.92\pm0.24) + (0.13\pm0.41) \: \logMstar  \\  \textrm{if} \ \logMstar > -0.68
    \end{split}
    \label{dM_acc_fitt}
\end{equation}
both relations are shown as light blue dashed lines in Figure \ref{fig:accretion_plots_medians}.

Table \ref{tab:information_criteria2} reports the values we evaluate for the three information criteria for this elected case.
Also in this case, the three information criteria prefer the broken power law over the linear model (even if with a weakened signal, in particular for $R^2$), while still keeping  $p<<0.05$, confirming that the change in the slope is real and can not be ascribed only to the presence of the sources with \Mstar $\gtrsim 0.7$. 

\begin{table}[!t]
    \centering
    \caption{Summary of the results for the selected information criteria adopted in this work to compare the single and broken power-law models obtain from our selected data (\logMstar $\lesssim -0.15$ ).}
    \begin{tabular}{c|c|c|c|c|c}
     \hline
     \hline
                    &  $R^2$ & AIC & BIC & p            & Case\\
     \hline
     \logLacc       &   0.60 & -311 & -293 & -          &single power-law\\
     \logLacc       &   0.61 & -325 & -308 & 5.8 $\times10^{-2}$      &broken power-law\\
     \hline
     \logdMacc      &   0.44 & -238 & -221 & -          &single power-law\\
     \logdMacc      &   0.48 & -273 & -256 & 5.5$\times10^{-5}$ &broken power-law\\
     \hline
    \end{tabular}
    \label{tab:information_criteria2}
\end{table}

\clearpage
\bibliography{biblio} 

\end{document}